\documentclass[11pt,a4paper,pdftex]{article}
\pdfoutput=1
\usepackage{jcappub}

\newcommand{\bi}{\bibitem}
\newcommand{\be}{\begin{eqnarray}}
\newcommand{\ee}{\end{eqnarray}}
\newcommand{\rar}{\rightarrow}

\title{Evolution of the spin parameter of accreting compact objects with non-Kerr quadrupole moment}

\author{Cosimo Bambi}

\affiliation{
Institute for the Physics and Mathematics of the Universe, 
The University of Tokyo, Kashiwa, Chiba 277-8583, Japan}

\emailAdd{cosimo.bambi@ipmu.jp}

\abstract{
There is robust observational evidence supporting the existence 
of $5 - 20$~$M_\odot$ compact bodies in X-ray binary systems
and of $10^5 - 10^9$~$M_\odot$ bodies at the center of many 
galaxies. All these objects are commonly interpreted as black
holes, even is there is no direct evidence that they have
an event horizon. A fundamental limit for a black hole in
4-dimensional general relativity is the Kerr bound $|a_*| \le 1$,
where $a_*$ is the spin parameter. This is just the condition 
for the existence of the event horizon. The accretion
process can spin a black hole up to $a_* \approx 0.998$
and some super-massive objects in galactic nuclei could be
rapidly rotating black holes with spin parameter close to 
this limit. However, if these super-massive objects are not
black holes, the Kerr bound does not hold and the accretion 
process can spin them up to $a_* > 1$. In this
paper, I consider compact bodies with non-Kerr quadrupole
moment. I study the evolution of the spin parameter due to
accretion and I find its equilibrium value. Future experiments
like the gravitational wave detector LISA will be able to
test if the super-massive objects at the center of galaxies
are the black holes predicted by general relativity. If
they are not black holes, some of them may be super-spinning
objects with $a_* > 1$.}

\begin{document}

\maketitle


\section{Introduction}

Today we have robust observational evidence for the existence of
$5 - 20$~$M_\odot$ compact objects in X-ray binary systems~\cite{bh1}
and of $10^5 - 10^9$~$M_\odot$ objects at the center of many 
galaxies~\cite{bh2}. The stellar-mass objects in X-ray binary
systems are surely too heavy to be neutron or quark stars for any
reasonable equation of state~\cite{rr,kal}. At least some of the
super-massive objects in galactic nuclei are too heavy and compact
to be clusters of non-luminous bodies, since the cluster lifetime
due to evaporation and physical collisions would be shorter than 
the age of the system~\cite{maoz}. All these objects are commonly 
interpreted as black holes (BHs), since they cannot be explained 
otherwise without introducing new physics. However, there is no
direct observation evidence that they have an event 
horizon~\cite{abra}, while there are theoretical arguments 
suggesting that the final product of the gravitational collapse may 
be quite different from a classical BH~\cite{np1,np2,np3,np4,np5,np6}.

In 4-dimensional general relativity, BHs are described by the Kerr
solution and are completely specified by two parameters: the mass,
$M$, and the spin angular momentum, $J$. Instead of $J$, one
can use the specific spin angular momentum $a=J/M$, or the
dimensionless spin parameter $a_*=J/M^2$. The fact that these 
objects have only two degrees of freedom is known as ``no-hair'' 
theorem~\cite{thm1,thm2,thm3} and implies that all the mass moments, 
${\cal M}_l$, and all the current moments, ${\cal S}_l$, of the 
space-time can be written in term of $M$ and $J$ by the following 
simple formula~\cite{mom1,mom2}:
\be
{\cal M}_l + i {\cal S}_l = M
\left(i\frac{J}{M}\right)^{\it{l}} \, .
\ee
As it was put forward by Ryan in~\cite{ry1,ry2}, by measuring the
mass, the spin, and at least one more non-trivial moment of 
the gravitational field of a BH candidate, one over-constrains 
the theory and can test the Kerr BH hypothesis.

The possibility of testing the Kerr metric around astrophysical
BH candidates with future experiments is quite extensively 
discussed in the recent literature. A constraint on the nature
of these objects has been recently obtained by considering the
mean radiative efficiency of AGN~\cite{agn}. The detection of 
gravitational waves from the inspiral of a stellar-mass compact 
body into a super-massive object, the so-called extreme mass 
ratio inspiral (EMRI), can be used to put very interesting
constraints~\cite{emri1,emri2,barack,emri3,emri4,emri5,emri6}. 
Since the future gravitational wave detector LISA will be able to 
observe about $10^4 - 10^6$ gravitational wave cycles emitted 
by an EMRI while the stellar-mass body is in the strong field 
region of the super-massive object, the quadrupole moment of the 
latter will be measured with a precision at the level of 
$10^{-2} - 10^{-4}$~\cite{barack}. The metric around super-massive 
BH candidates can also be probed by observing their 
``shadow''~\cite{sha1,sha2,sha3}. Additional proposals to test 
the Kerr BH hypothesis involve the possible discovery of a 
stellar-mass BH candidate with a radio pulsar as companion~\cite{wex}, 
the study of the K$\alpha$ iron lines~\cite{ps}, and the analysis 
of the X-ray spectrum of a geometrically thin and optically 
thick accretion disk~\cite{noi}.

A fundamental limit for a BH in general relativity is the Kerr
bound $|a_*| \le 1$. This is just the condition for the 
existence of the event horizon. The accretion process can spin 
a BH up to $a_* \approx 0.998$~\cite{thorne} and many 
super-massive objects in galactic nuclei may be
rapidly rotating BHs with spin parameter close to this 
limit~\cite{m1,m2,ho}. Nevertheless, if these objects are not 
the BH predicted by general relativity, the Kerr bound does not 
hold. Interestingly, in this case the accretion process can easily
spin them up to $a_* > 1$~\cite{ss}. Such a possibility is 
currently ignored in the literature. It is also neglected in
those works in which it is not assumed that these objects are
BHs and it is studied how future observations can test the 
Kerr metric. For experiments like LISA, that rely on matched 
filtering, this may be a serious problem.

In Ref.~\cite{sim1,sim2,sim3,sim4}, I studied some features of 
the accretion process onto objects with $|a_*| > 1$. However, an 
important question to address is if objects with $|a_*| > 1$ can 
really form. In~\cite{ss}, 
I showed that deviations from the Kerr metric can
have the accreting gas spin the body up to $a_* > 1$. I used the 
Manko--Novikov (MN) metric~\cite{mn}, which is a stationary, 
axisymmetric, and asymptotically flat exact solution of the vacuum
Einstein's equation. It describes the exterior gravitational 
field of a rotating massive body with arbitrary mass multipole 
moments and has an infinite number of free parameters. However, 
it has the drawback that it is valid only for sub-extreme objects, 
with spin parameter $|a_*| < 1$. So, I showed that the equilibrium
spin parameter $a_*^{eq}$ must be larger than 1, but it was impossible 
to discuss the accretion process for $a_*>1$, compute $a_*^{eq}$,
and figure out the properties of the space-time when $a_*>1$. 
An extension to include super-spinning objects, if it exists, is 
non-trivial. In the present paper, I overcome this problem by 
considering another metric, the Manko--Mielke--Sanabria-G\'omez 
(MMS) solution~\cite{mms1,mms2}. It is not as general as the MN 
metric, but it can be easily extended to discuss objects with $a_* > 1$.

The content of this paper is as follows. In Sec.~\ref{s-2}, I
present the MMS solution adapted to the case of super-spinning objects.
In Sec.~\ref{s-3}, I study the main properties of this space-time
and I show that the evolution of the spin parameter for an object
whose exterior gravitational field is described by the MMS metric
can be calculated as in the case
of a BH. In Sec.~\ref{s-4}, I discuss the evolution of
the spin parameter of an object with non-Kerr quadrupole moment,
as a consequence of the accretion process. Sec.~\ref{s-5} is 
devoted to the discussion of the results presented in Sec.~\ref{s-4},
while summary and conclusions are in Sec.~\ref{s-6}. Throughout 
the paper I use units in which $G_N = c = 1$.

\section{Exterior field of a rotating body with non-Kerr 
quadrupole moment \label{s-2}}

The Manko--Mielke--Sanabria-G\'omez (MMS) metric~\cite{mms1,mms2}
is a stationary, axisymmetric, reflection-symmetric, and 
asymptotically flat exact solution of the Einstein--Maxwell 
equations. It includes, as special cases, the Kerr and the
$\delta=2$ Tomimatsu--Sato solutions. It is specified by five 
real parameters: the mass, $M$, the specific spin angular momentum, 
$a = J/M$, the electric charge, $\mathcal{Q}$, and two other 
parameters, $b$ and $\mu$. The latter of which determine the mass
quadrupole moment, $Q$, and the magnetic dipole moment, $\mathcal{M}$. 
Since the MMS solution is reflection-symmetric, all the odd 
mass moments and the even current moments are identically zero. 
Here I am interested in the vacuum solution only, and I put 
$\mathcal{Q}=\mathcal{M}=0$, which implies $\mu=0$. The 
mass quadrupole moment of the gravitational field is 
\be
Q = Q_{\rm Kerr} - (d - \delta - ab)M \, ,
\ee
where $Q_{\rm Kerr} = - a^2 M$ is the quadrupole moment of a BH
and
\be
\delta = -\frac{M^2 b^2}{M^2 - (a-b)^2} \, , \qquad
d = \frac{M^2 - (a-b)^2}{4} \, .
\ee

In Refs.~\cite{mms1,mms2}, the metric is written in prolate
spheroidal coordinates, which are suitable for slow-rotating objects. 
It can be adapted to fast-rotating objects by proceeding as for 
the $\delta=2$ Tomimatsu--Sato metric~\cite{ccs}: one has to 
change the prolate spheroidal coordinates into oblate spheroidal
coordinates. That can be achieved through the transformation:
\be
x \rar ix \, , \quad k \rar -ik \, ,
\ee
where $i$ is the imaginary unit, i.e. $i^2 = -1$.
The line element becomes
\be
ds^2 &=& - f \left(dt - \omega d\phi\right)^2
+ \frac{k^2 e^{2\gamma}}{f}\left(x^2 + y^2\right)
\left(\frac{dx^2}{x^2 + 1} + \frac{dy^2}{1 - y^2}\right) 
+ \nonumber\\ &&
+ \frac{k^2}{f} \left(x^2 + 1\right)\left(1 - y^2\right) d\phi^2 \, ,
\ee
where $k = \sqrt{-d-\delta}$ and
\be
f = \frac{A}{B} \, , \quad
\omega = - (1 - y^2) \frac{C}{A} \, , \quad
e^{2\gamma} = \frac{A}{16 k^8 (x^2 + y^2)^4} \, .
\ee
The functions $A$, $B$, and $C$ can be written in the following 
compact way~\cite{mms2}
\be
A = R^2 + \lambda_1 \lambda_2 S^2 \, , \quad
B = A + R P + \lambda_2 S T \, , \quad
C = R T - \lambda_1 S P \, .
\ee
Here $\lambda_1 = k^2 (x^2 + 1)$, $\lambda_2 = y^2 - 1$, and
\be
P &=& 2kMx[(2kx + M)^2 - 2y^2(2\delta + ab - b^2) - a^2 + b^2]
- 4y^2(4\delta d - M^2 b^2) \, , \nonumber\\
R &=& 4[k^2(x^2+1)+\delta(1-y^2)]^2+(a-b)
[(a-b)(d-\delta) - M^2b](1-y^2)^2\, , \nonumber\\
S &=& -4\{(a-b)[k^2(x^2+y^2)+2\delta y^2]+y^2M^2b\} 
\, , \nonumber\\
T &=& 8Mb(kx + M)[k^2(x^2 + 1) + \delta(1 - y^2)] +
\nonumber\\ &&
+ (1 - y^2)\{(a - b)(M^2b^2 - 4\delta d)
- 2M(2kx + M)[(a-b)(d-\delta) - M^2 b]\} \, .
\ee

\begin{figure}
\par
\begin{center}
\includegraphics[type=pdf,ext=.pdf,read=.pdf,width=7cm]{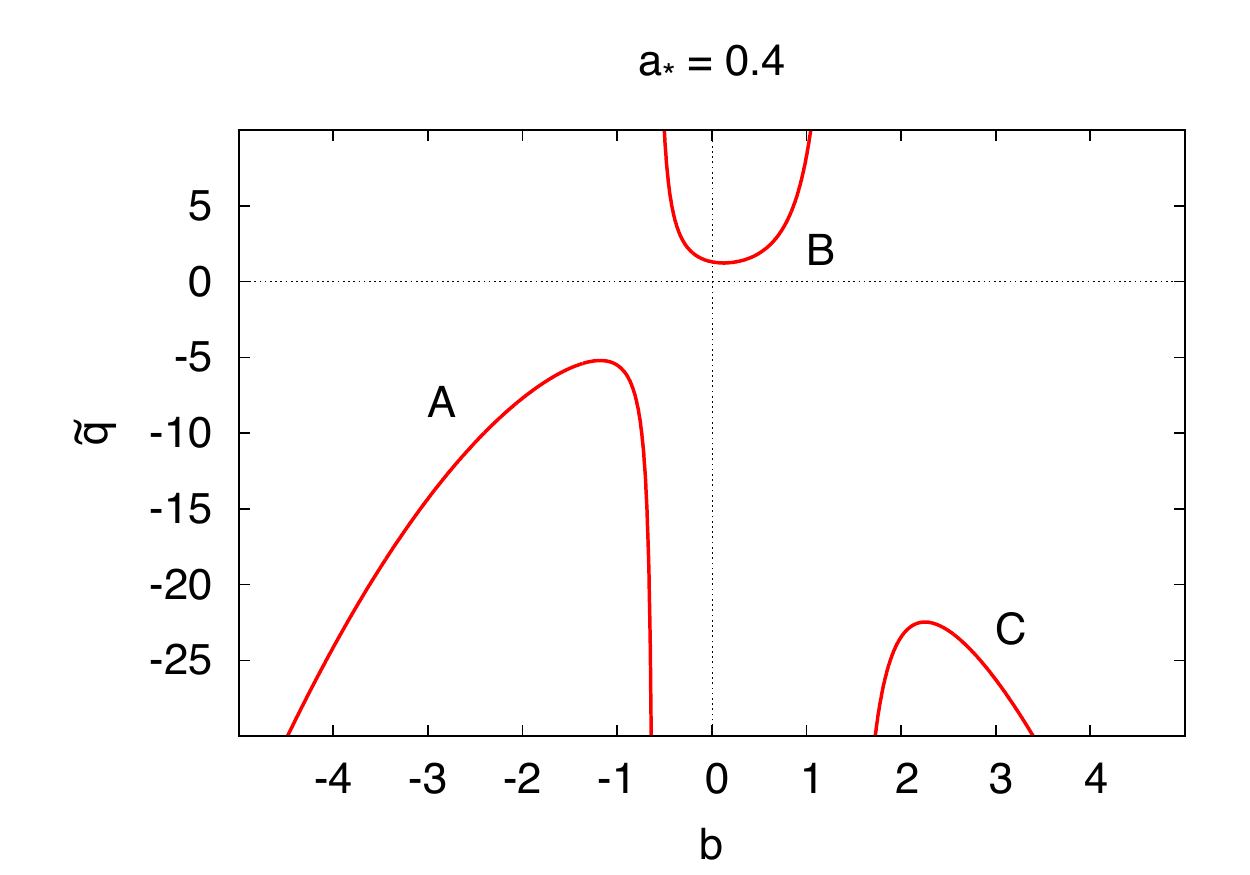}
\includegraphics[type=pdf,ext=.pdf,read=.pdf,width=7cm]{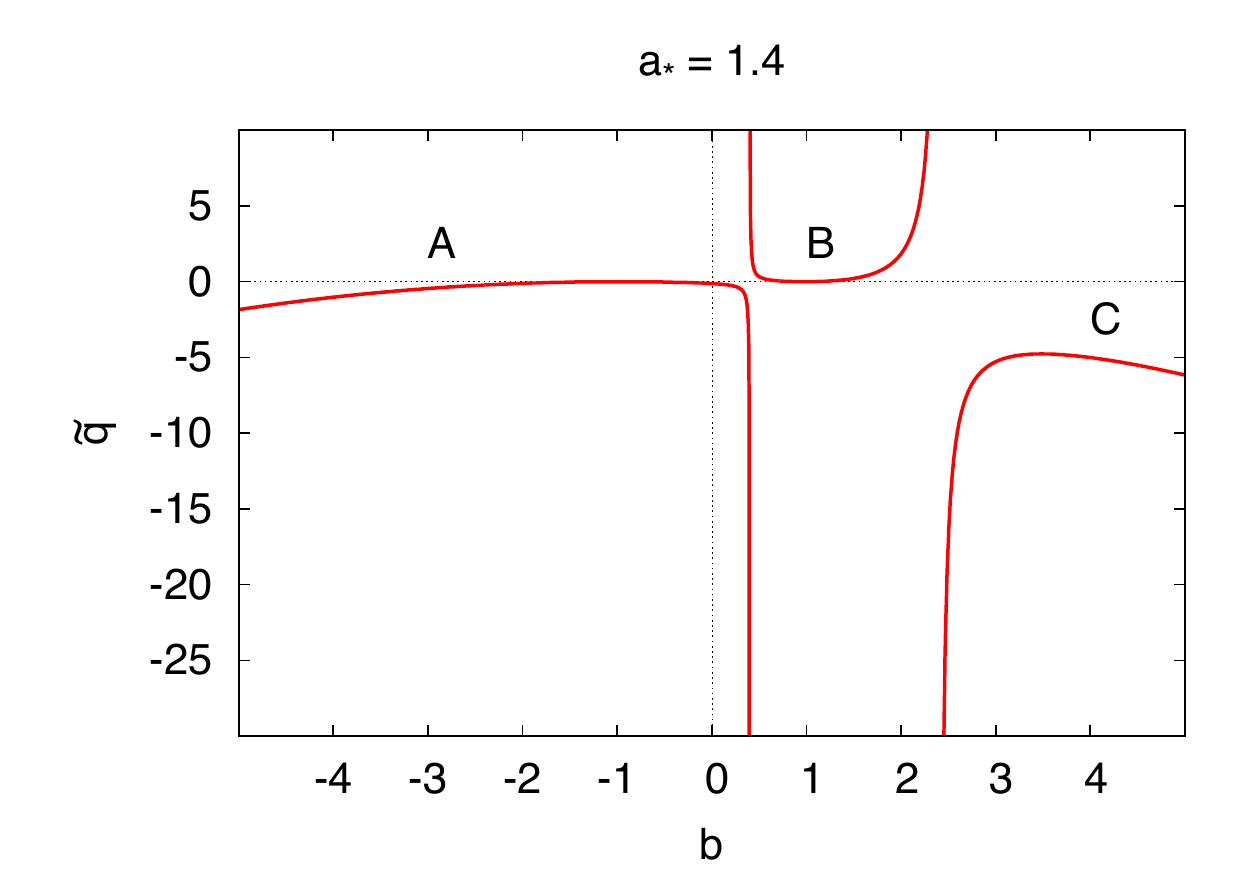}
\end{center}
\par
\vspace{-5mm} 
\caption{$\tilde{q}=(d-\delta-ab)/a^2$ as a function of the parameter
$b$ for $a_*=0.4$ (left panel) and $a_*=1.4$ (right panel). 
The Kerr solution is recovered for $b = \pm \sqrt{a^2 - M^2}$. 
$b$ is given in units of $M=1$.}
\label{f-2-1}
\end{figure}

In general, the mass quadrupole moment of an object depends on 
its mass and on its spin angular momentum in a non-trivial way, 
according to the specific properties of the matter the body is 
made of. In this paper, I will use the following definition 
of anomalous quadrupole moment $\tilde{q}$
\be\label{eq-q}
Q = - (1 + \tilde{q}) a^2 M \, .
\ee
For $\tilde{q} = 0$, one finds $Q=Q_{\rm Kerr}$, while for 
$\tilde{q}>0$ ($\tilde{q}<0$) the object is more oblate 
(prolate) than a BH. Eq.~(\ref{eq-q}) is what we expect for 
neutron stars, with $\tilde{q} \approx 1 - 10$, depending on the 
equation of state and the mass of the body~\cite{neutron}. It 
is reasonable to assume that $\tilde{q}\ge-1$, since otherwise 
the effect of the rotation would make the body more and more 
prolate. For given $M$ and $a$, the MMS solution does not allow 
for any arbitrary value of the anomalous quadrupole moment. 
In Fig.~\ref{f-2-1}, I show $\tilde{q} = (d - \delta - ab)/a^2$
as a function of $b$ for $a_*=0.4$ (left panel)
and $a_*=1.4$ (right panel). There are three distinct curves:
one in the region $b<a-M$ (curve A), one in the region
$a-M<b<a+M$ (curve B), and the last one for $b>a+M$ (curve C).
The equation $\tilde{q} = (d - \delta - ab)/a^2$ may have no
real solutions for $b$ (when $|a_*|<1$), and up to four distinct 
real solutions (for $\tilde{q}$ sufficiently small). 
The Kerr metric is recovered for $b^2 = a^2 - M^2$.
Two or more distinct solutions with the same $M$, $a$, and $\tilde{q}$
correspond to objects with the same mass, spin, and quadrupole 
moment, but with different higher order moments. 
A particular solution requires prolate spheroidal coordinates
if $d+\delta>0$ and oblate spheroidal coordinates if $d+\delta<0$;
otherwise, the constant $k$ becomes an imaginary number. Generally
speaking, slow-rotating solutions require prolate spheroidal 
coordinates, while fast-rotating solutions need oblate spheroidal
coordinates; however, the critical value $a_*^{crit}$ separating 
the two cases depends on $\tilde{q}$ and can be either larger or 
smaller than 1. In the rest of the paper, I will study the
evolution of the spin parameter of objects with $\tilde{q}\ge-1$ 
and therefore I will discuss the solutions of $b$ belonging to 
the curve A and B. When I find two solution with the same 
quadrupole moment, I will call the solution with smaller 
$|b|$, MMS1, and the other, MMS2.

\section{Properties of the MMS solution \label{s-3}}

\subsection{Structure of the space-time}

The structure of the MMS space-time reminds one of the
Tomimatsu--Sato and MN solutions~\cite{kodama, emri3}. One
finds naked singularities and closed time-like curves very
close to the massive object. The idea is that all these
pathological features do not exist because inside the
compact object, where the MMS metric, which is a vacuum 
solution, does not hold. Moreover, for $|a_*|<a_*^{eq}$
these pathological regions are always inside the inner
radius of the disk and therefore they do not play any
role in the discussion of the accretion process.

\begin{figure}
\par
\begin{center}
\includegraphics[type=pdf,ext=.pdf,read=.pdf,width=5cm]{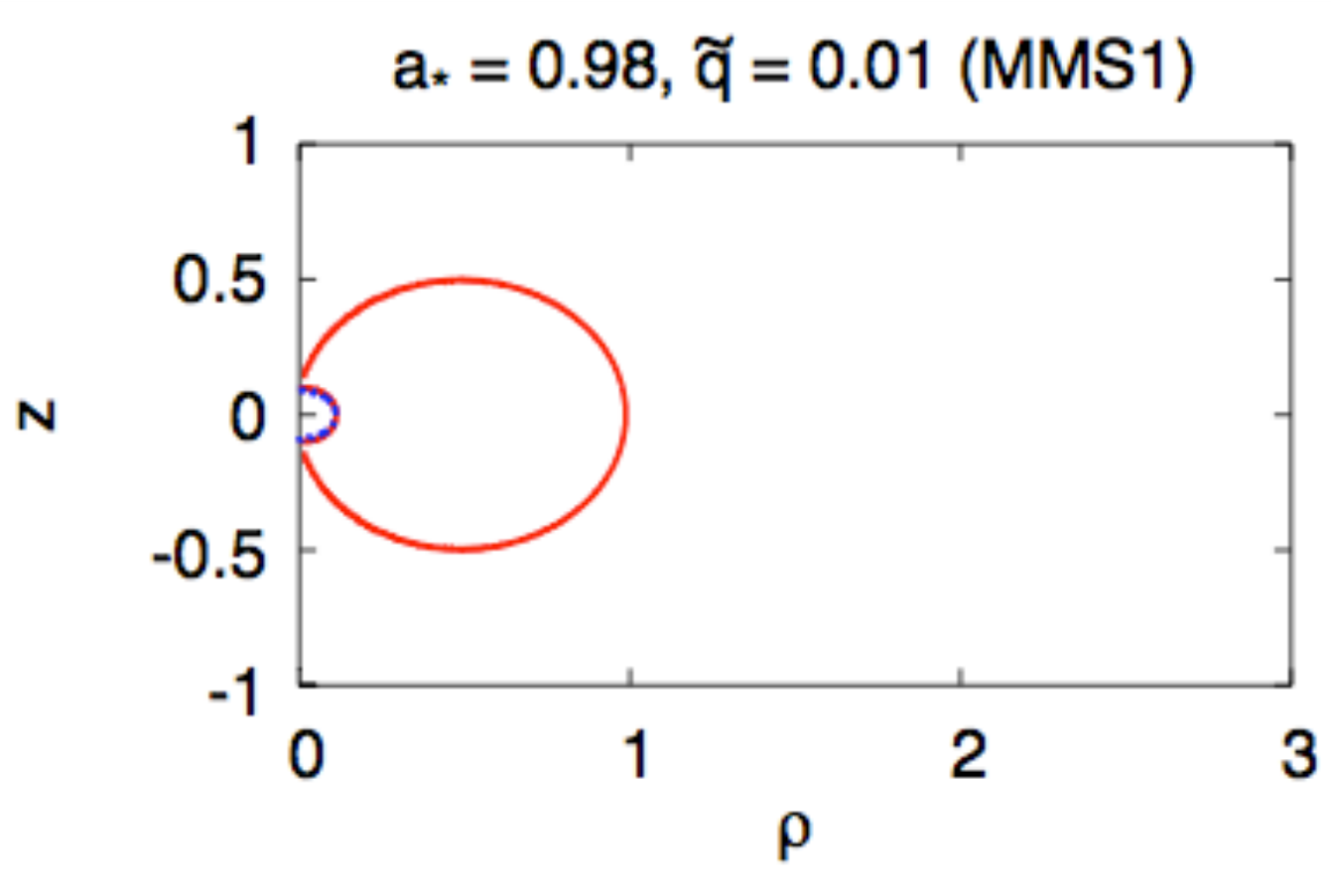}
\hspace{0.8cm}
\includegraphics[type=pdf,ext=.pdf,read=.pdf,width=5cm]{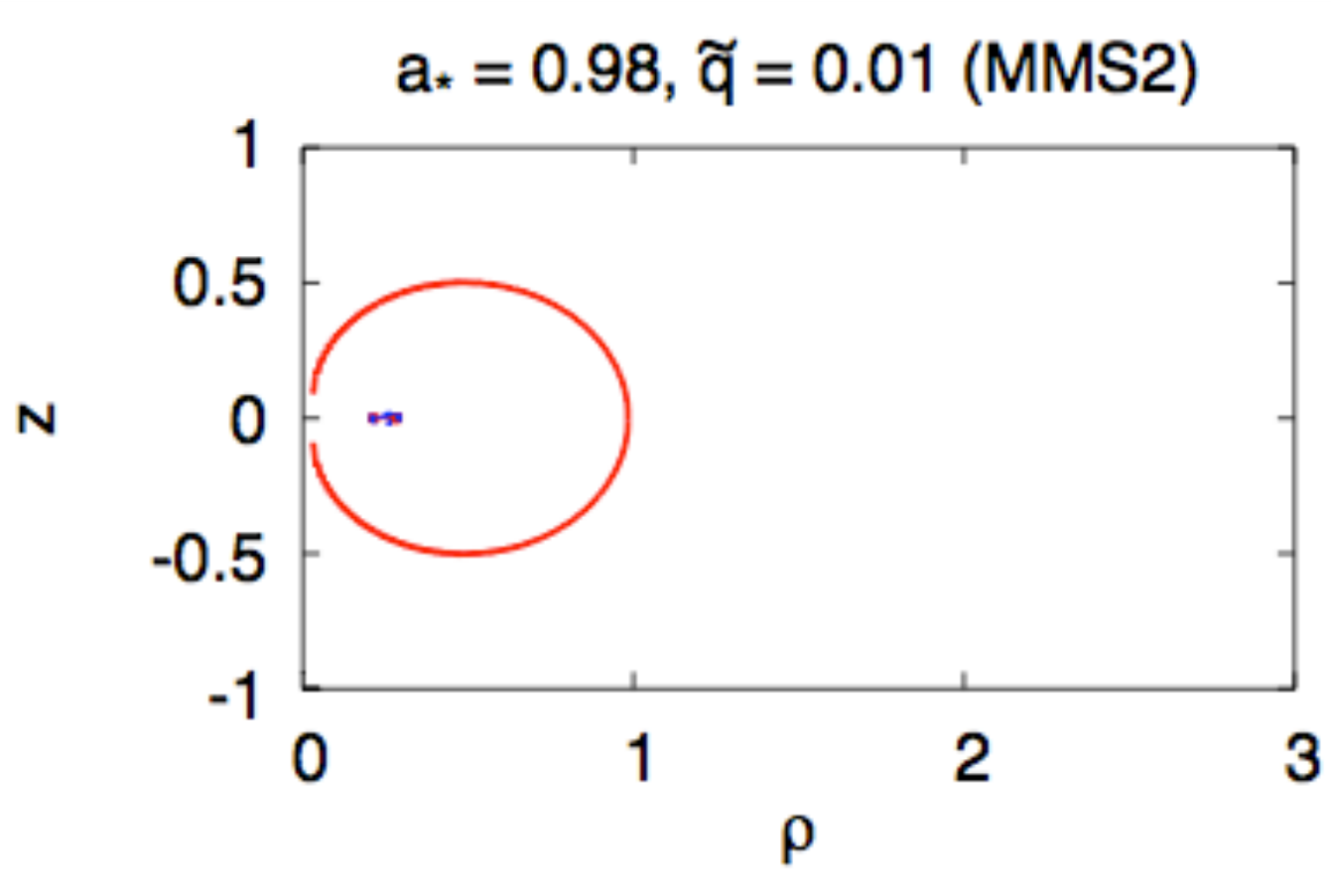} \\
\includegraphics[type=pdf,ext=.pdf,read=.pdf,width=5cm]{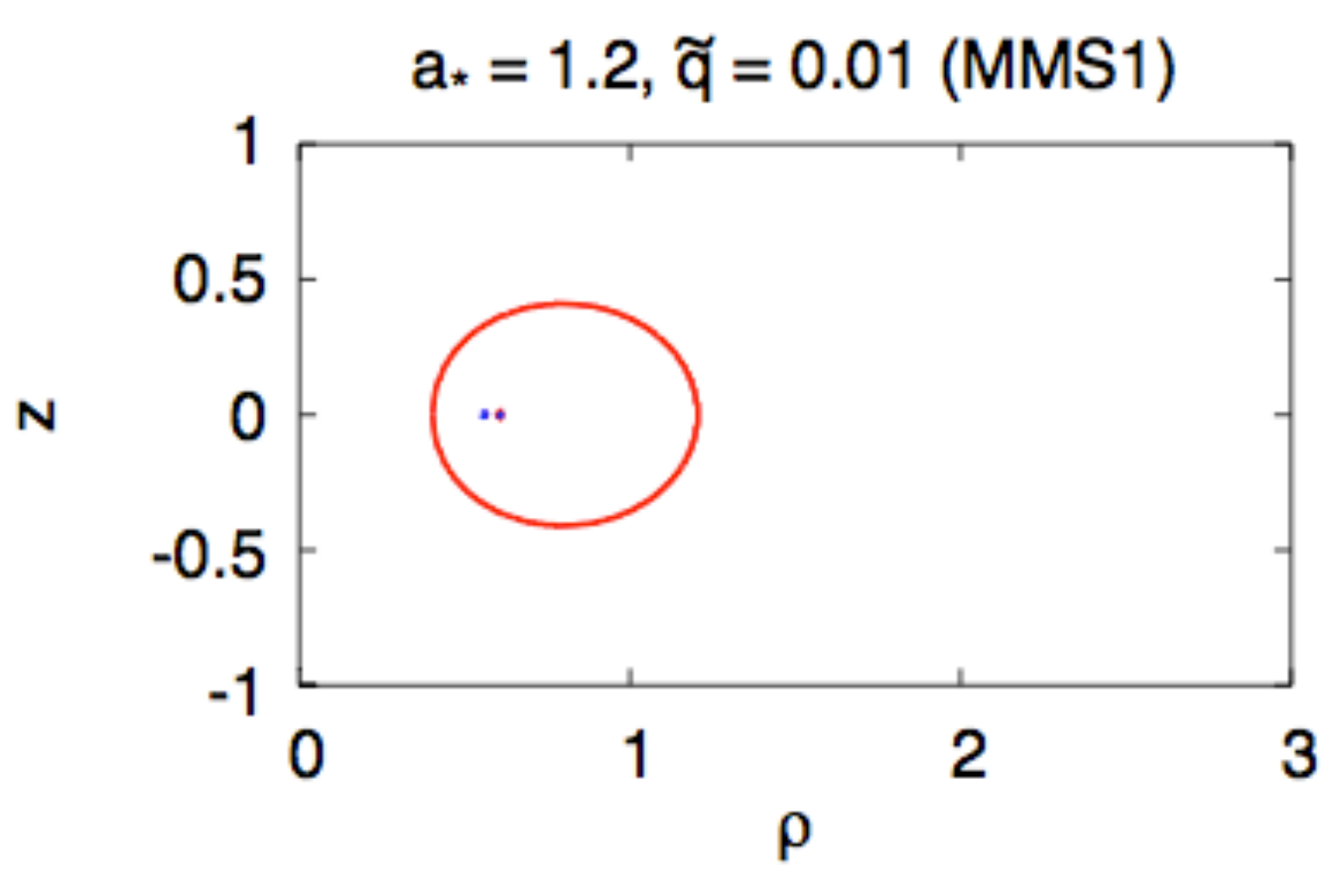}
\hspace{0.8cm}
\includegraphics[type=pdf,ext=.pdf,read=.pdf,width=5cm]{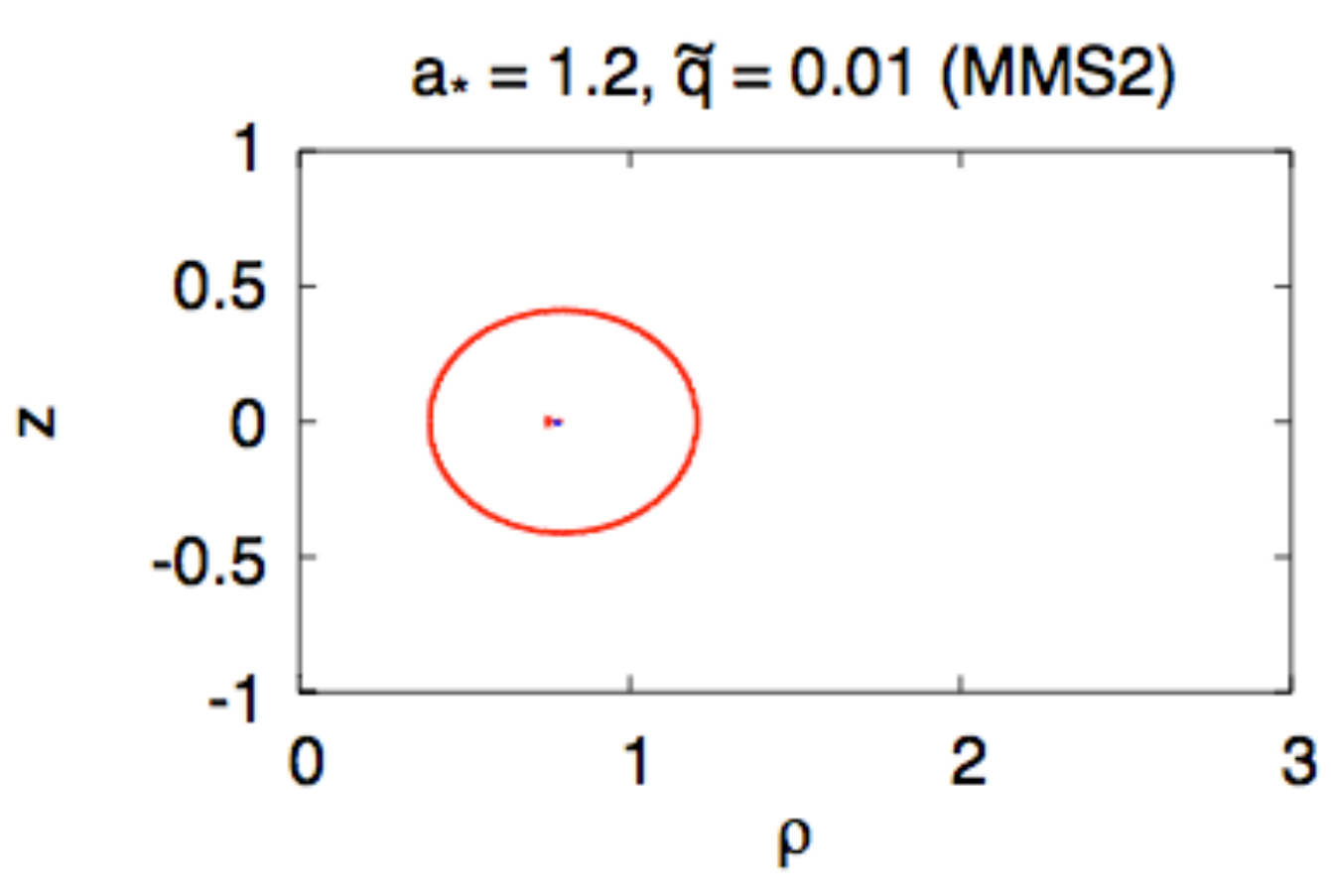}
\end{center}
\par
\vspace{-5mm} 
\caption{Space-time structure of the MMS solution for 
$\tilde{q} = 0.01$ and spin parameter $a_* = 0.98$ (top panels) and 
$a_* = 1.2$ (bottom panels). The solid red curves denote the 
infinite redshift surface $g_{tt} = 0$, defining the boundary 
of the ergoregion of the space-time. The dotted blue curves 
define the boundary of the causality violating region, where 
$g_{\phi\phi} < 0$. $\rho$ and $z$ are given in units of $M=1$.}
\label{f-3-1}
\end{figure}

\begin{figure}
\par
\begin{center}
\includegraphics[type=pdf,ext=.pdf,read=.pdf,width=5cm]{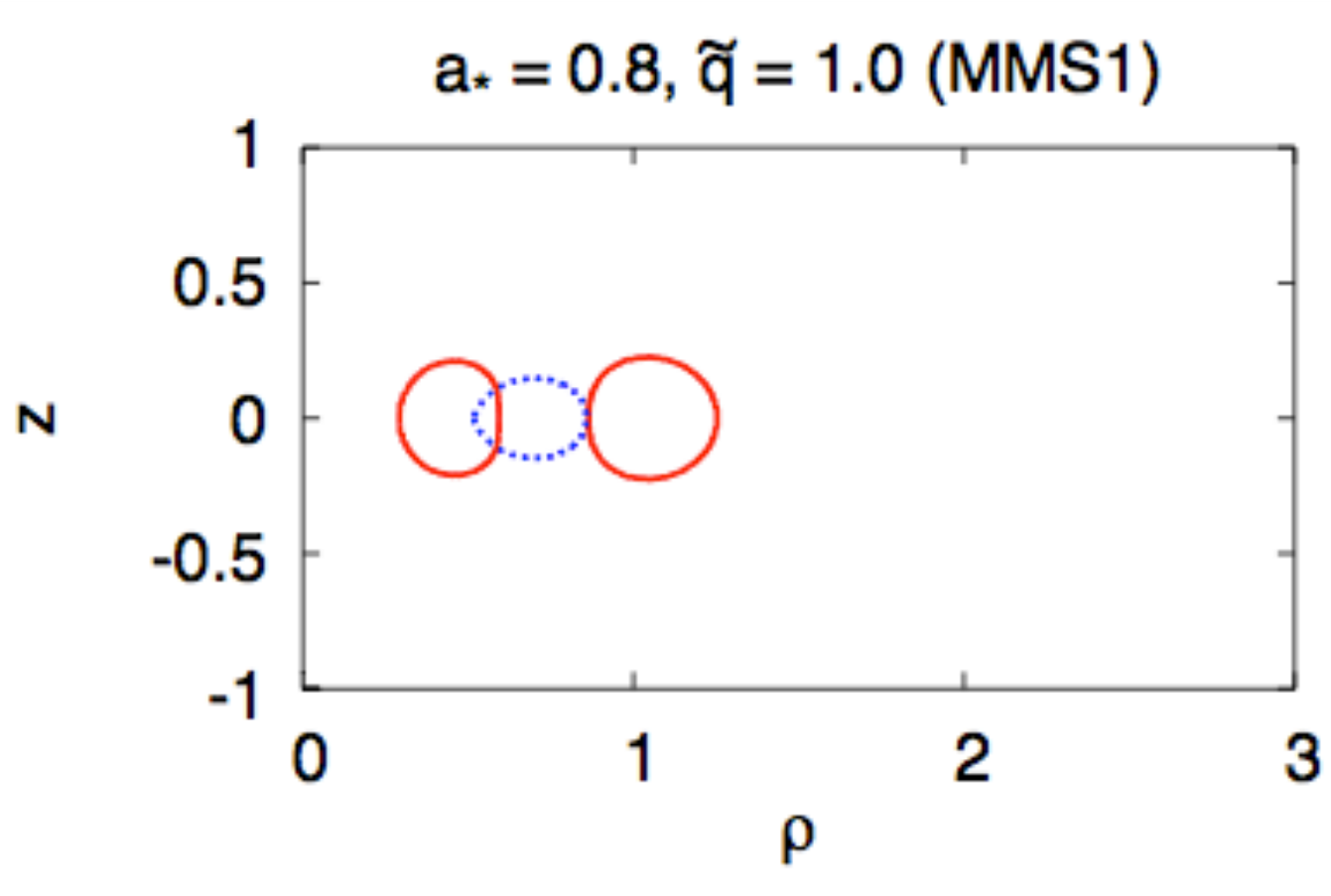}
\hspace{0.8cm}
\includegraphics[type=pdf,ext=.pdf,read=.pdf,width=5cm]{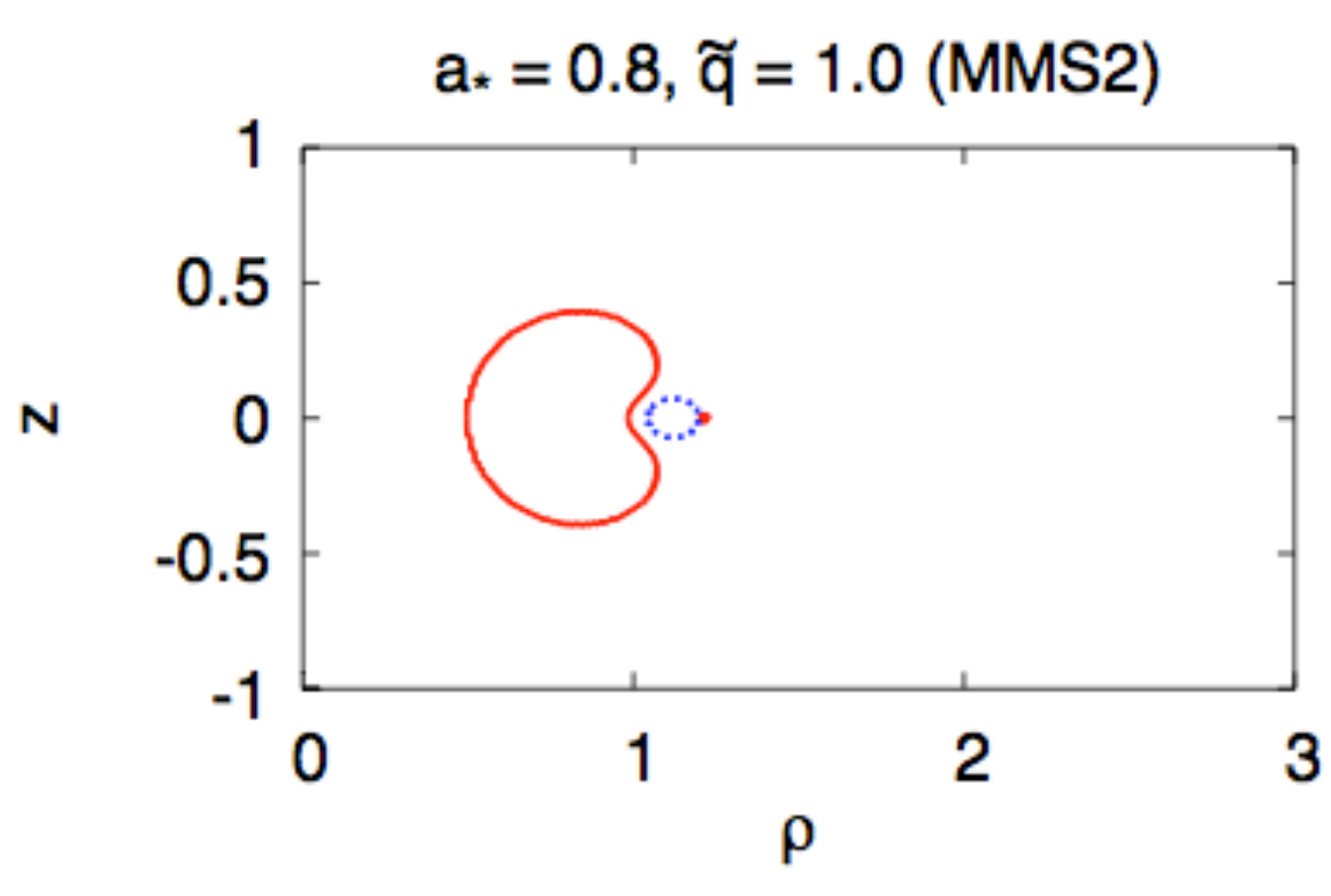} \\
\includegraphics[type=pdf,ext=.pdf,read=.pdf,width=5cm]{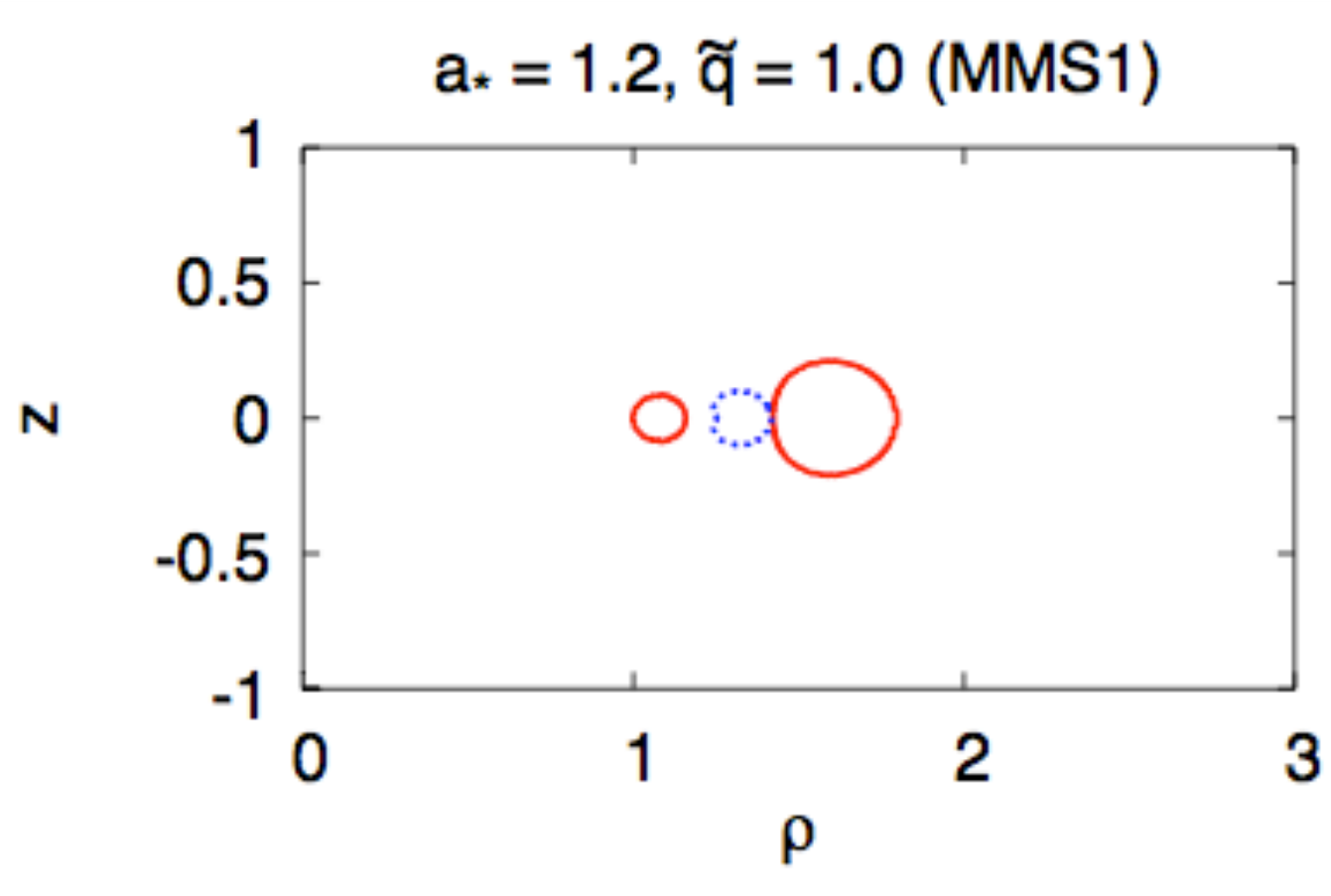}
\hspace{0.8cm}
\includegraphics[type=pdf,ext=.pdf,read=.pdf,width=5cm]{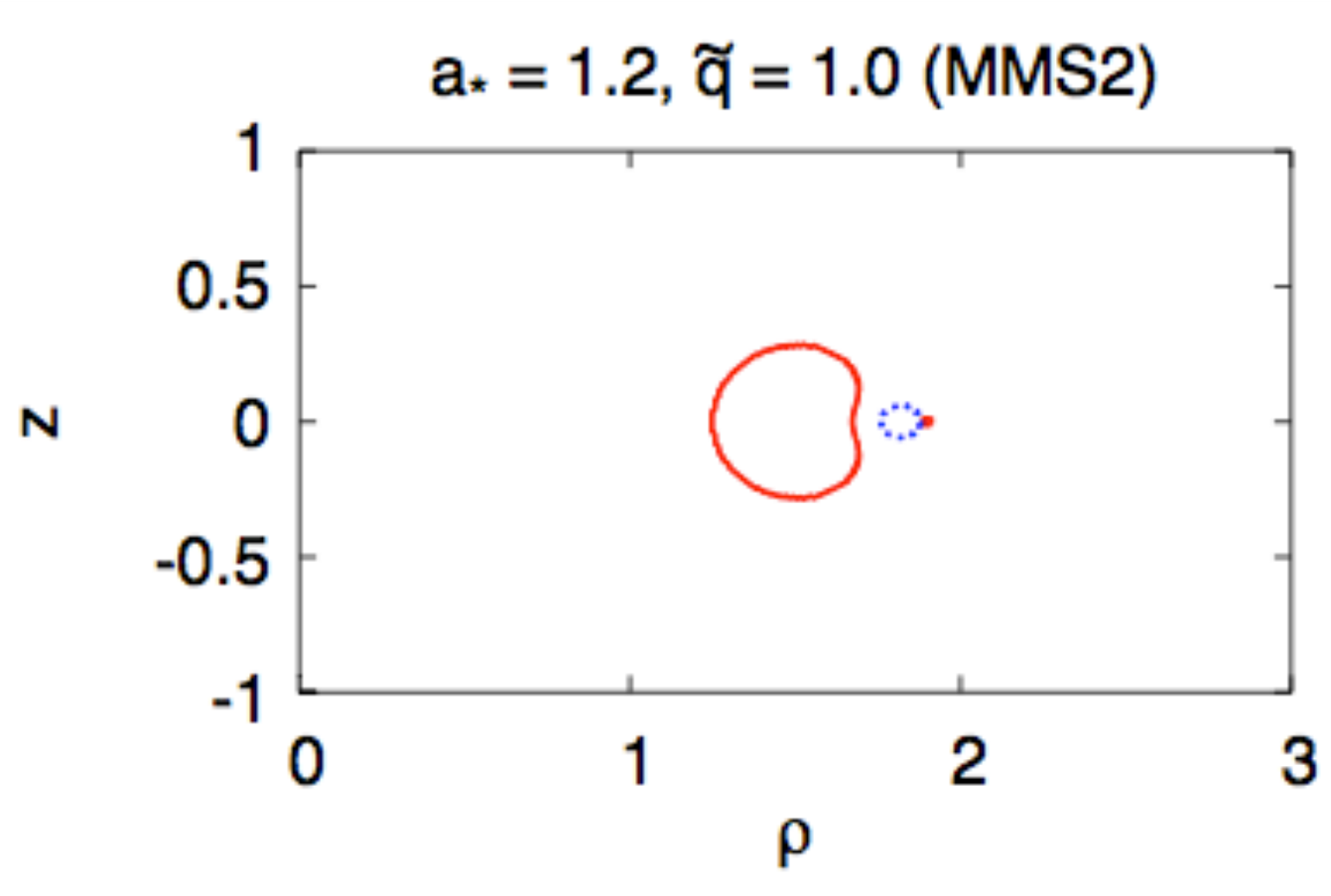}
\end{center}
\par
\vspace{-5mm} 
\caption{As in Fig.~\ref{f-3-1}, for $\tilde{q} = 1.0$ and spin 
parameter $a_* = 0.8$ (top panels) and $a_* = 1.2$ (bottom panels).}
\label{f-3-1b}
\end{figure}

\begin{figure}
\par
\begin{center}
\includegraphics[type=pdf,ext=.pdf,read=.pdf,width=5cm]{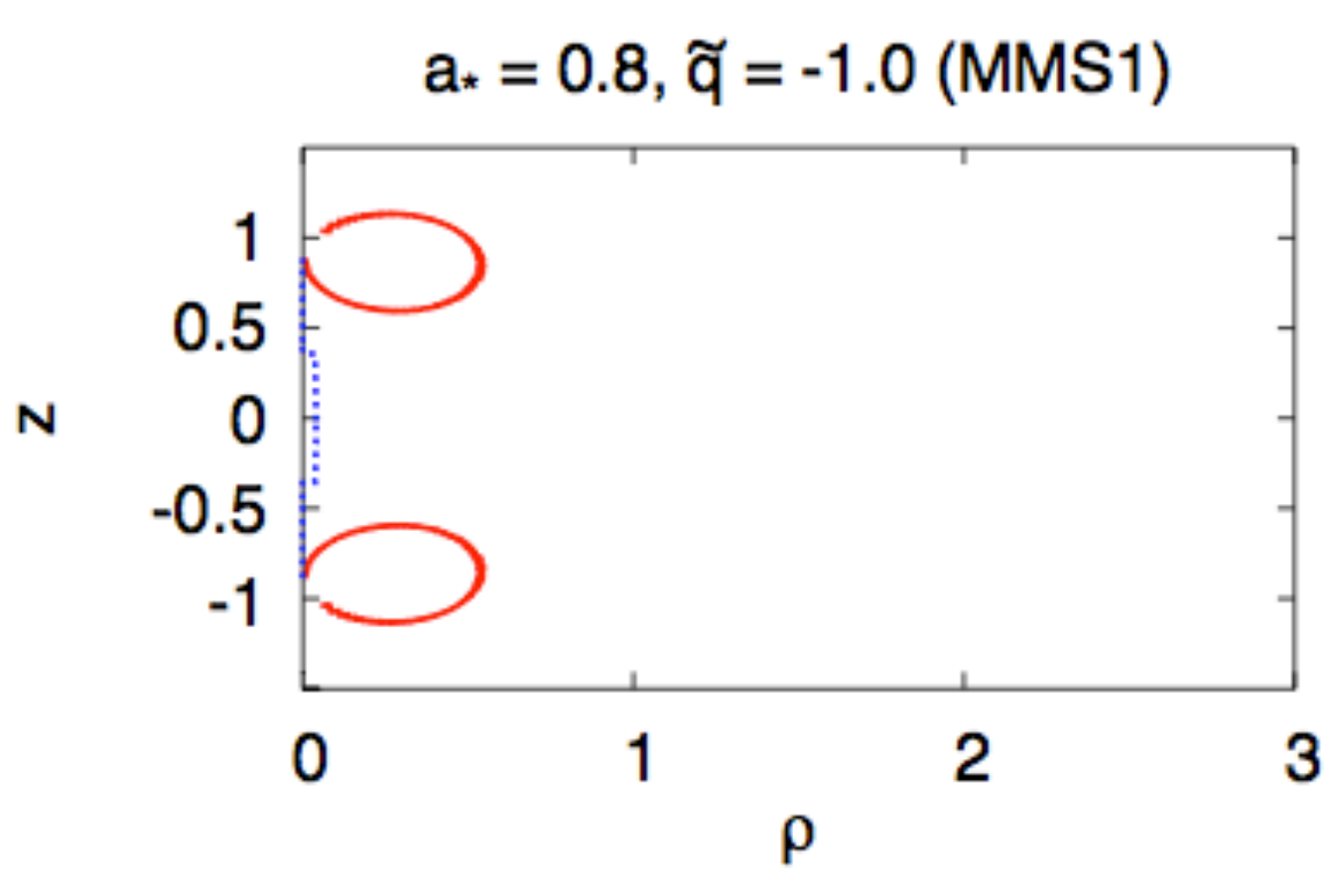}
\hspace{0.8cm}
\includegraphics[type=pdf,ext=.pdf,read=.pdf,width=5cm]{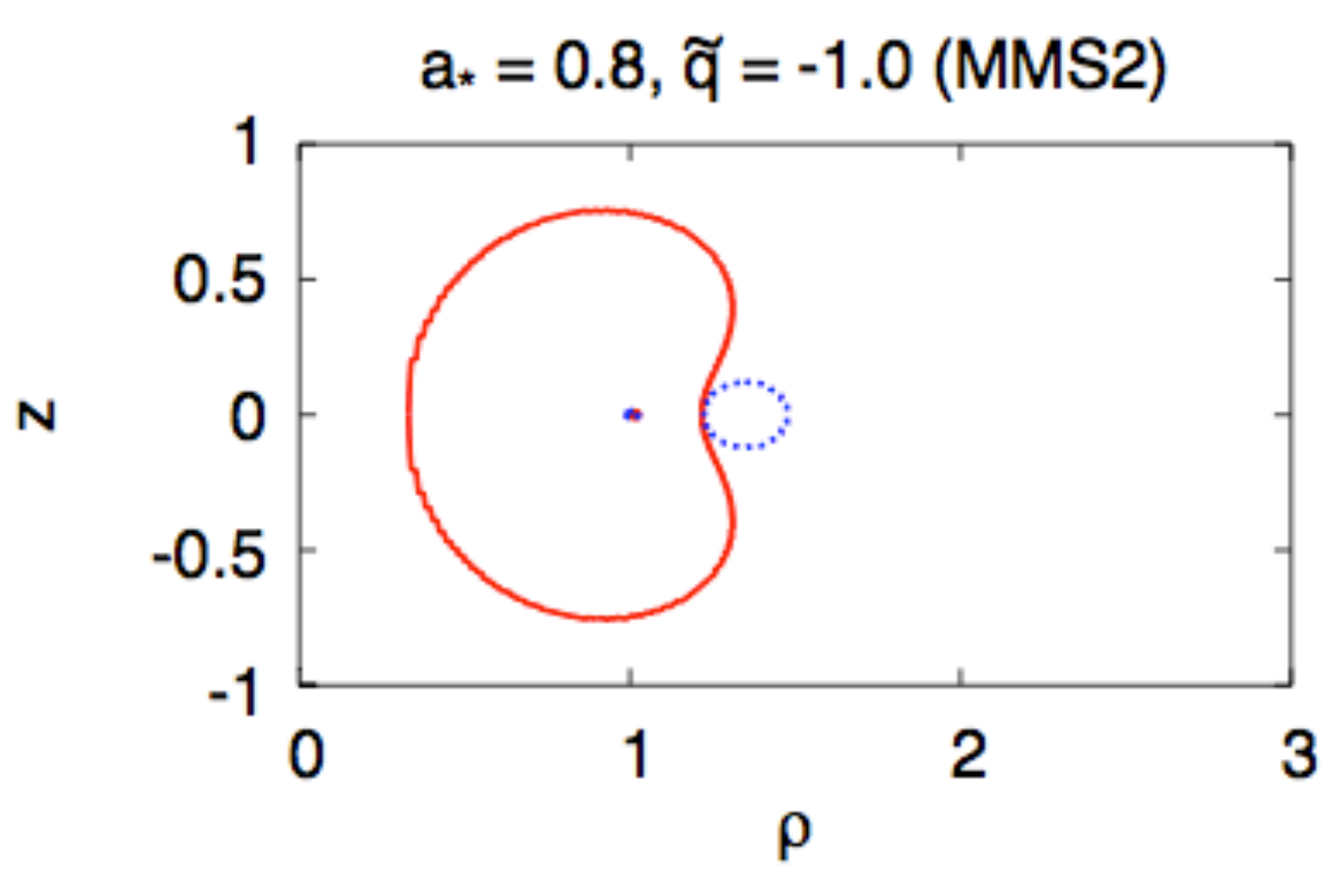} \\
\includegraphics[type=pdf,ext=.pdf,read=.pdf,width=5cm]{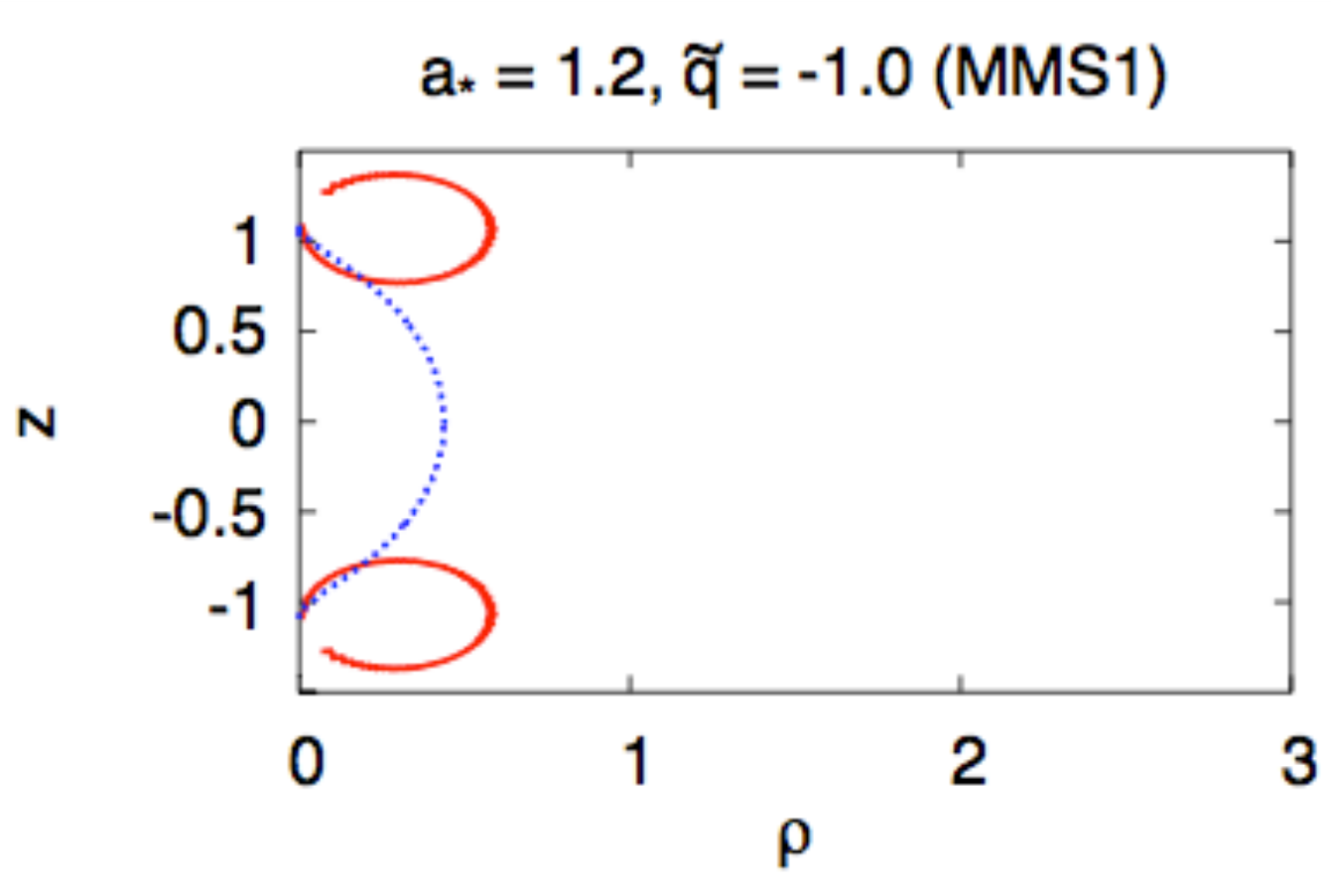}
\hspace{0.8cm}
\includegraphics[type=pdf,ext=.pdf,read=.pdf,width=5cm]{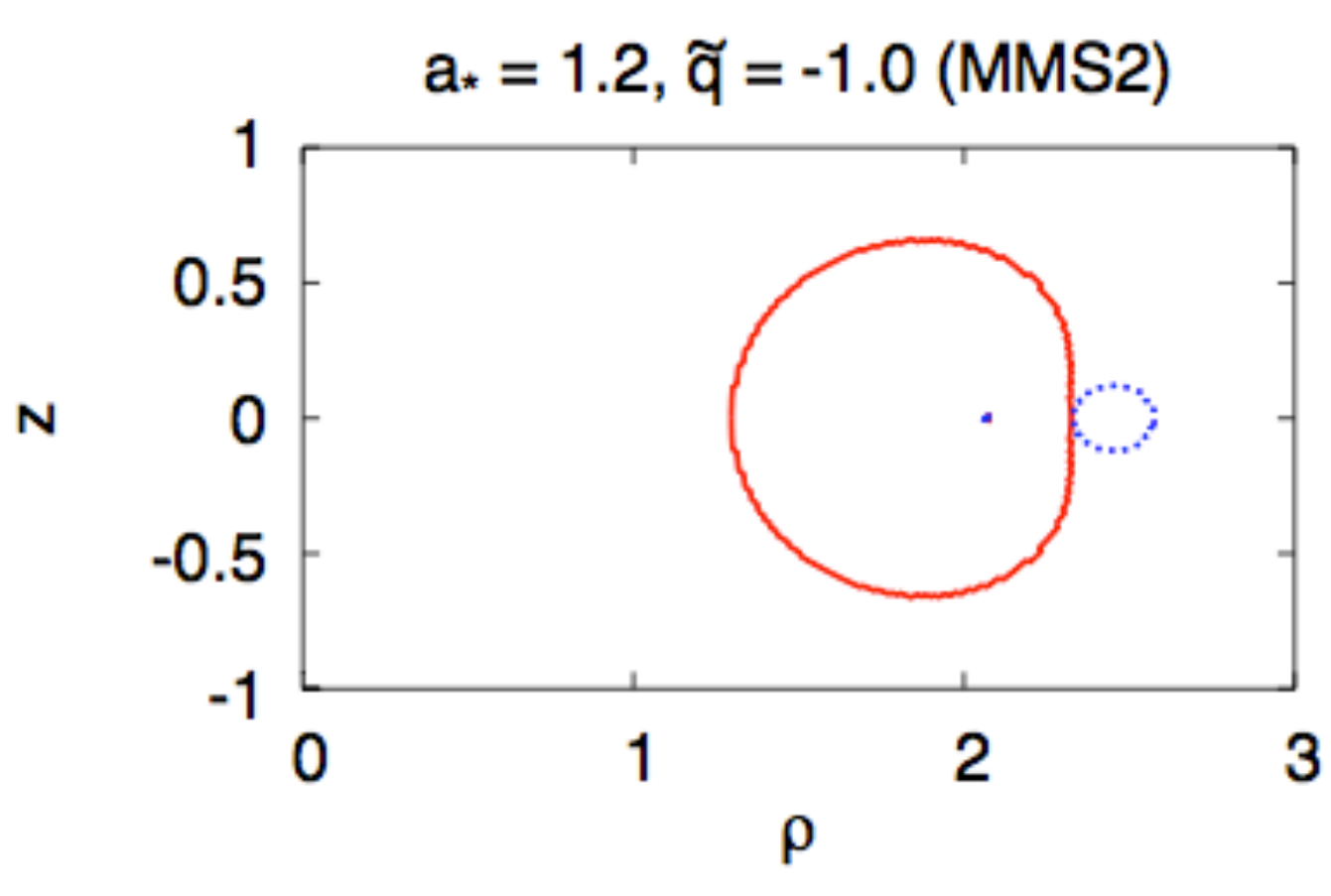}
\end{center}
\par
\vspace{-5mm} 
\caption{As in Fig.~\ref{f-3-1b}, for $\tilde{q} = -1.0$.}
\label{f-3-1c}
\end{figure}

The infinite redshift surface $g_{tt} = 0$ determines the
boundary of the ergoregion, which is not a pathological region
and exists even around a rotating BH. Figs.~\ref{f-3-1}, \ref{f-3-1b}, 
and \ref{f-3-1c} show the ergosphere (red solid curve) for some MMS
solutions with different value of the spin parameter and of
the anomalous quadrupole moment. I use quasi-cylindrical 
coordinates $\rho z$, see App.~\ref{a-1}. The topology of 
the ergoregion can change significantly as $a_*$ and $\tilde{q}$
vary. The same figures show also the boundary of the
closed time-like curve regions (blue dotted curves), defined
by $g_{\phi\phi} = 0$. For small deviations from the Kerr
metric (Fig.~\ref{f-3-1}, $\tilde{q}=0.01$), the shape of the 
ergoregion is very similar to the Kerr case, see App.~\ref{a-2}, 
but there are one or two small holes, where $g_{tt}<0$. Every 
hole of the ergoregion can be associated with a small region with
closed time-like curves, even if the boundaries of the holes
and of the closed time-like curve regions do not coincide.
For $\tilde{q}=1.0$ (Fig.~\ref{f-3-1b}), there are two 
disconnected ergoregions\footnote{The fact that there are two
disconnected ergoregions may depend on the coordinate system,
as in prolate spheroidal coordinates the Schwarzschild radius
$r = M$ reduces to the segment $|\rho| < \sqrt{a^2 - M^2}$ and 
$z = 0$ and the region with $r < M$ is not included, see 
App.~\ref{a-1}. The existence of these ergoregions should also
be taken with caution, because close to the pathological region 
with closed time-like curves.} with no holes and one closed time-like 
curve region. For $\tilde{q}=-1.0$ (Fig.~\ref{f-3-1c}), one 
finds two ergoregions and one closed time-like curve region
for the MMS1 solutions (left panels) and one ergoregion with a
hole and two distinct closed time-like curve regions for the
MMS2 solutions (right panels). Let us notice that the MMS1
solutions with $(a_*,\,\tilde{q})=\{(0.98,\,0.01),(0.8,\,1.0),
(1.2,\,1.0)\}$ require prolate spheroidal coordinates, while 
all the other cases need oblate spheroidal coordinates. 
As the spin parameter 
increases, ergoregions and closed time-like curve regions move to 
larger $\rho$, but this is just an artifact of the coordinate
system, see App.~\ref{a-1}.

\subsection{Geodesic motion} 

\begin{figure}
\par
\begin{center}
\includegraphics[type=pdf,ext=.pdf,read=.pdf,width=5cm]{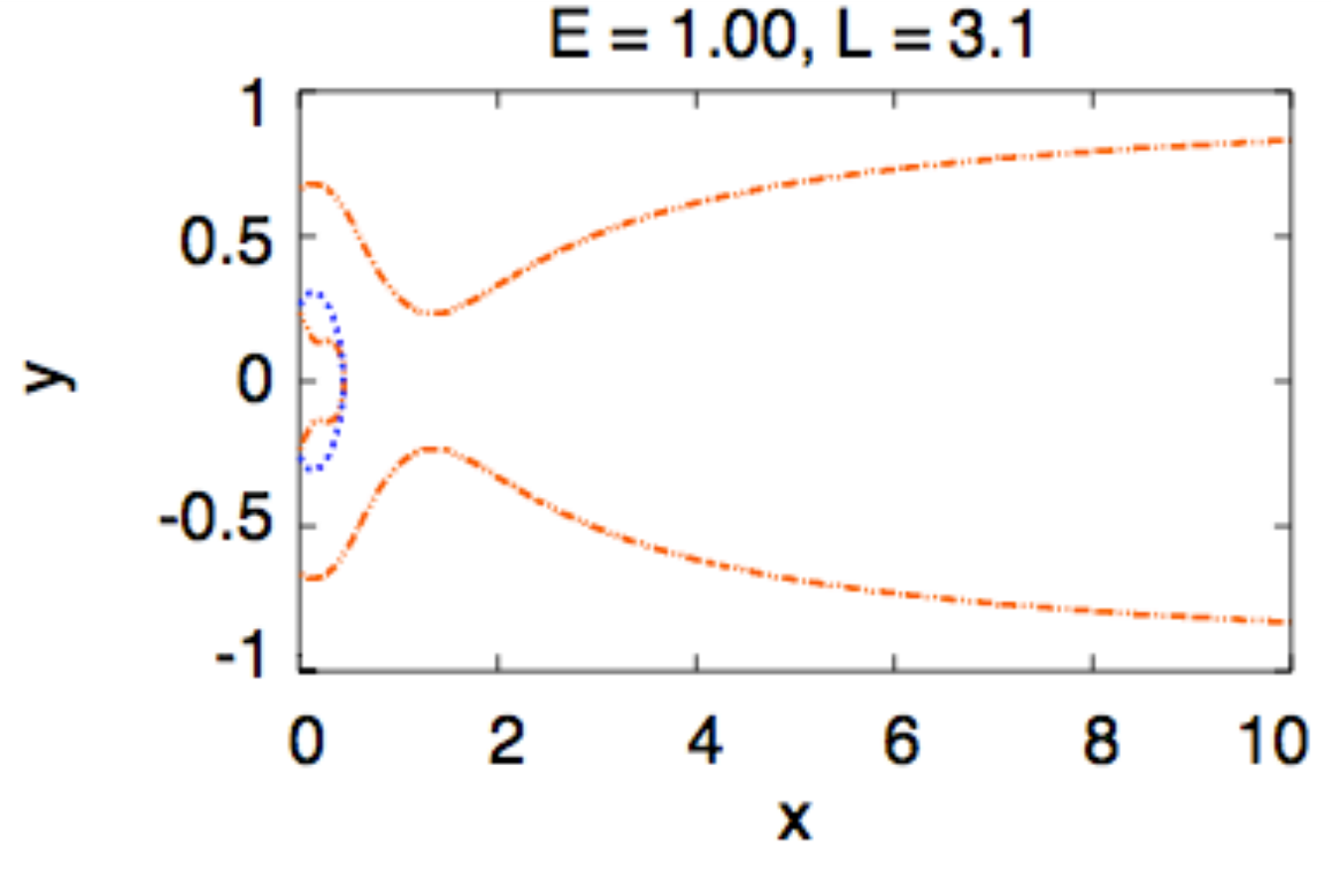}
\includegraphics[type=pdf,ext=.pdf,read=.pdf,width=5cm]{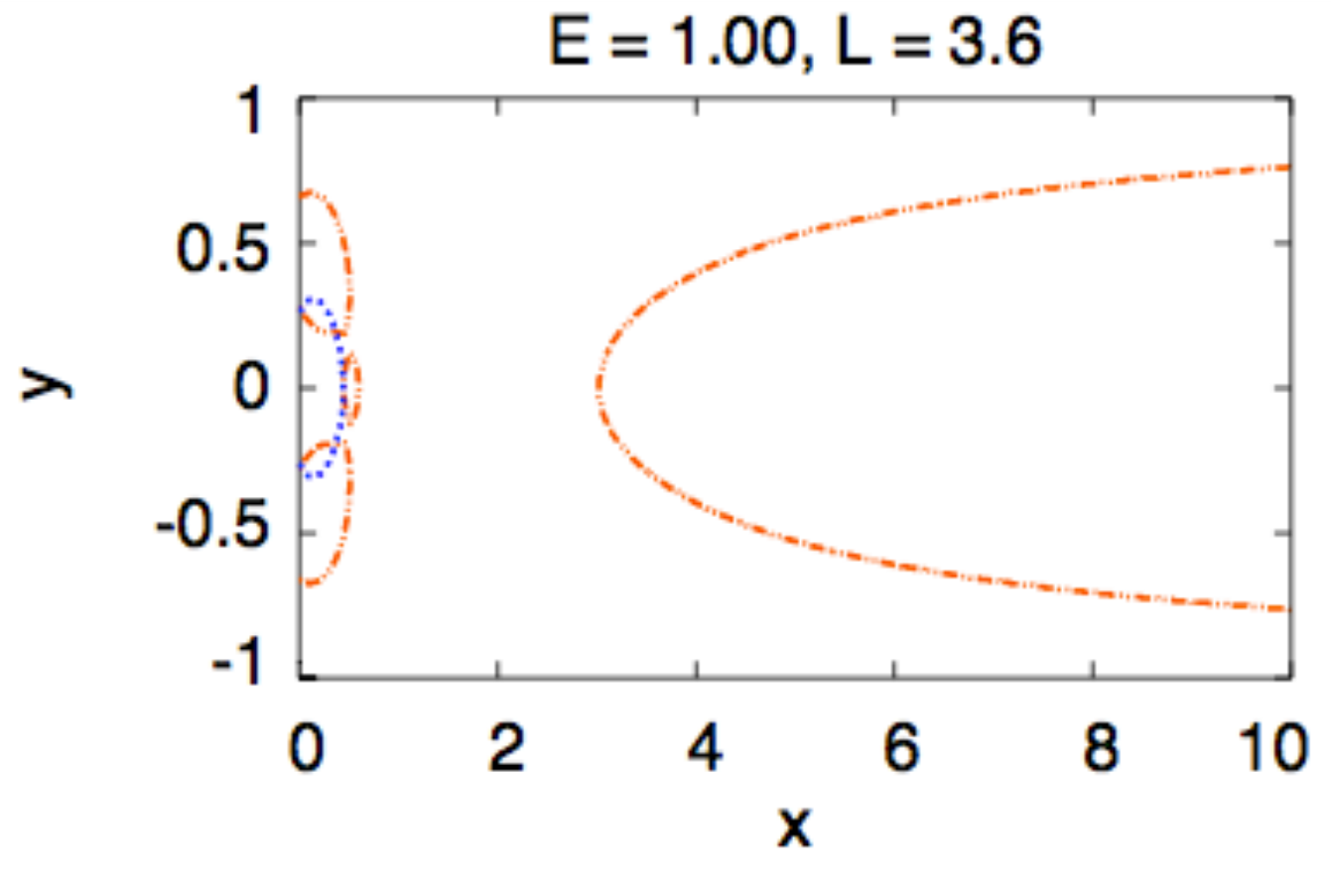}
\includegraphics[type=pdf,ext=.pdf,read=.pdf,width=5cm]{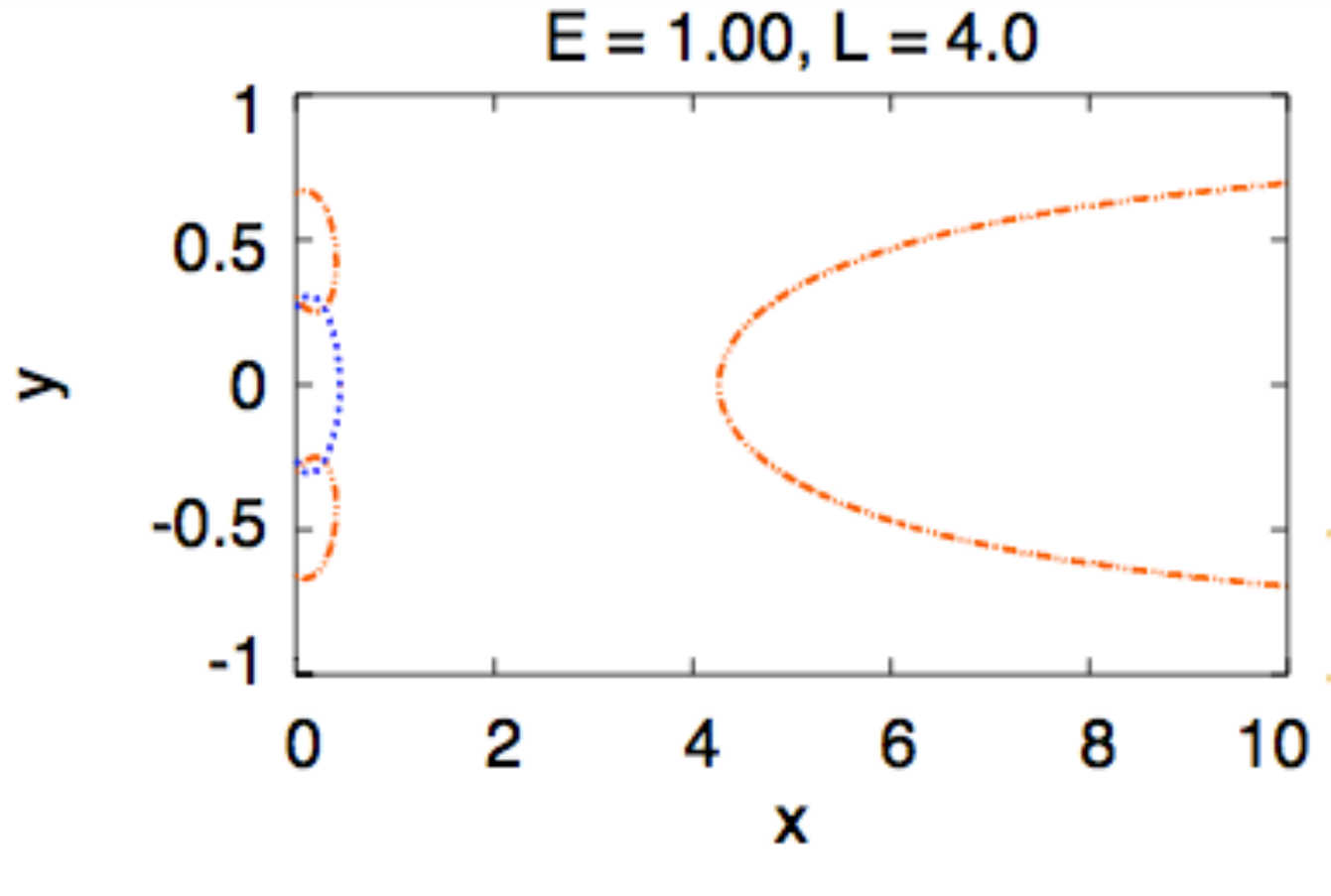} \\
\includegraphics[type=pdf,ext=.pdf,read=.pdf,width=5cm]{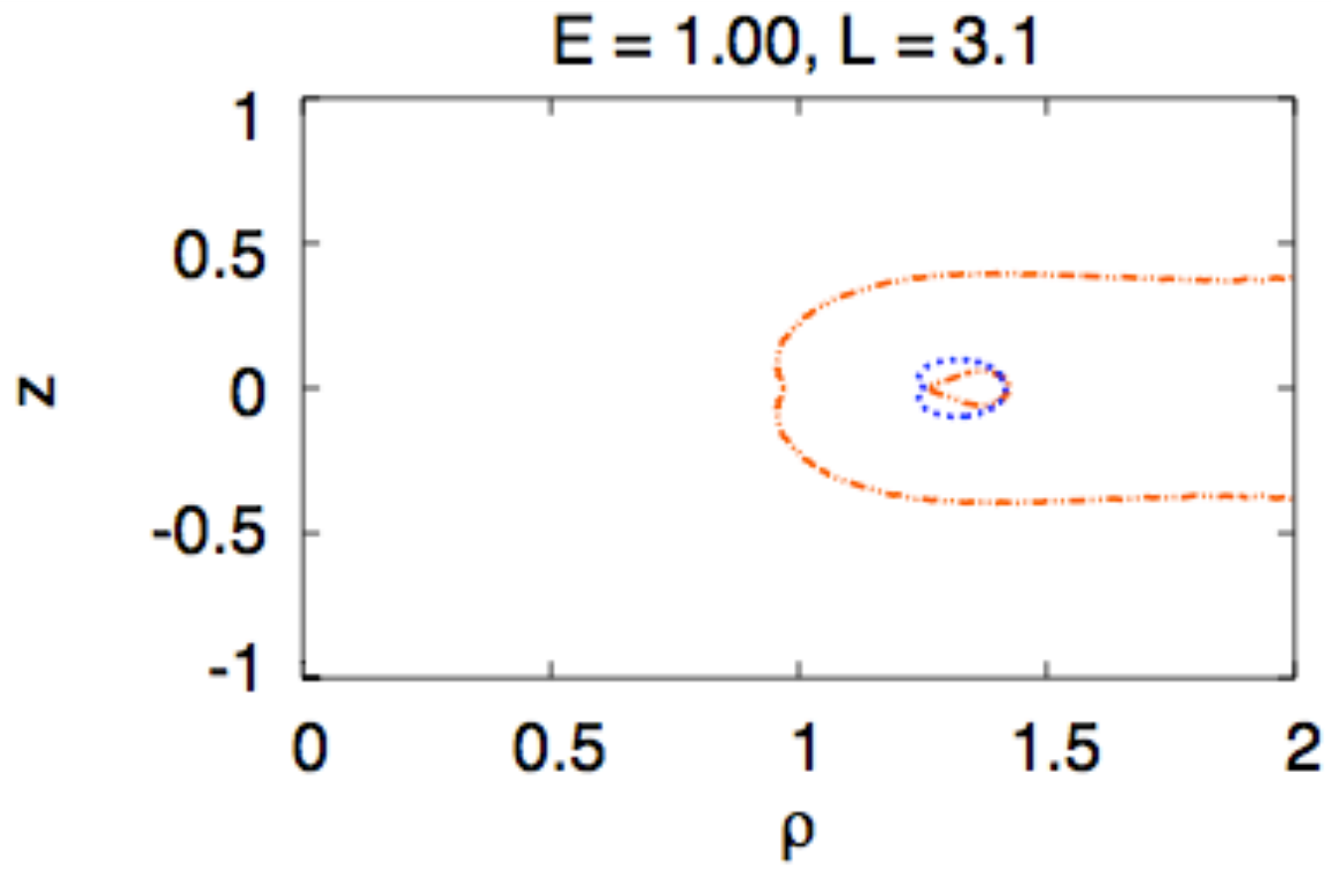}
\includegraphics[type=pdf,ext=.pdf,read=.pdf,width=5cm]{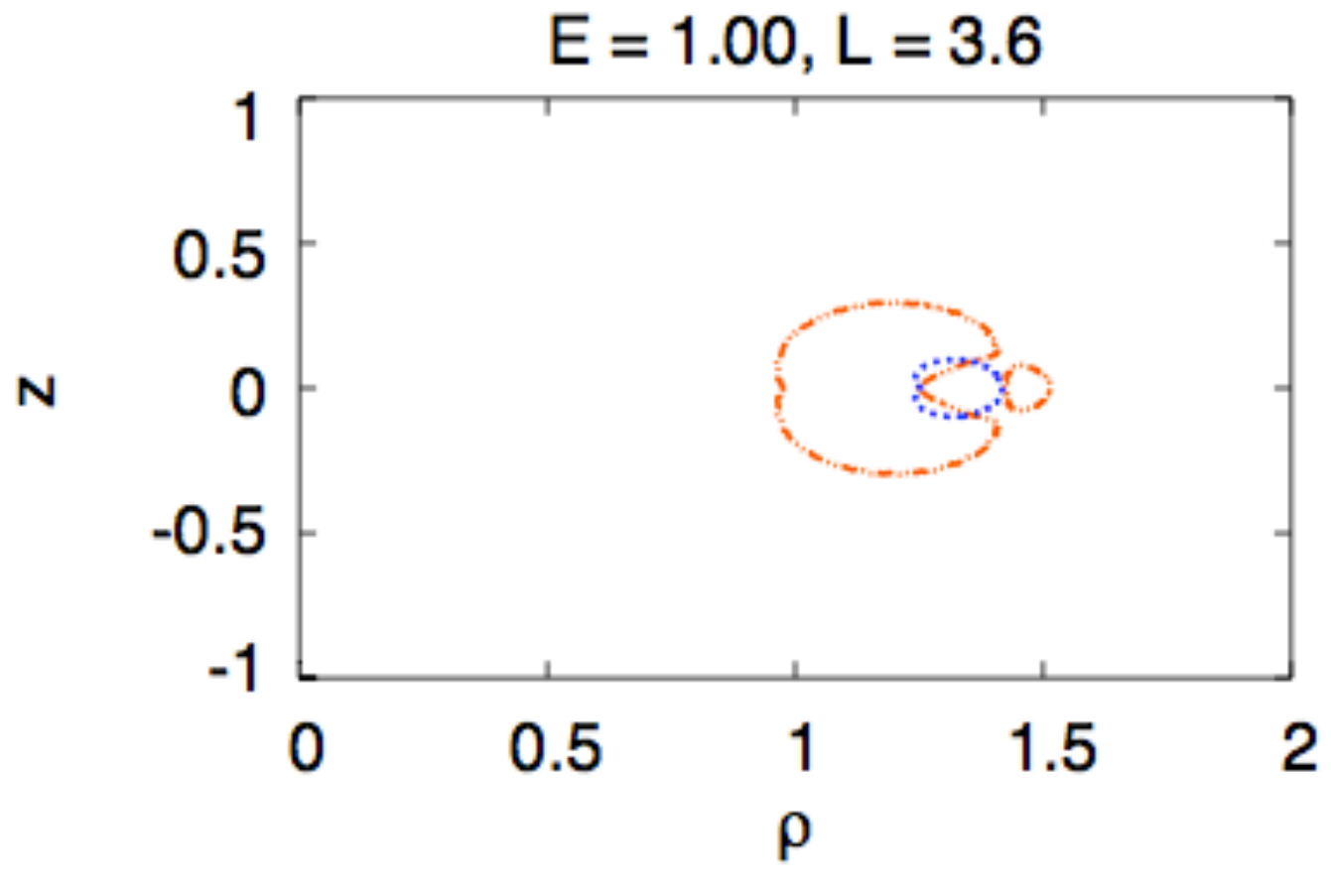}
\includegraphics[type=pdf,ext=.pdf,read=.pdf,width=5cm]{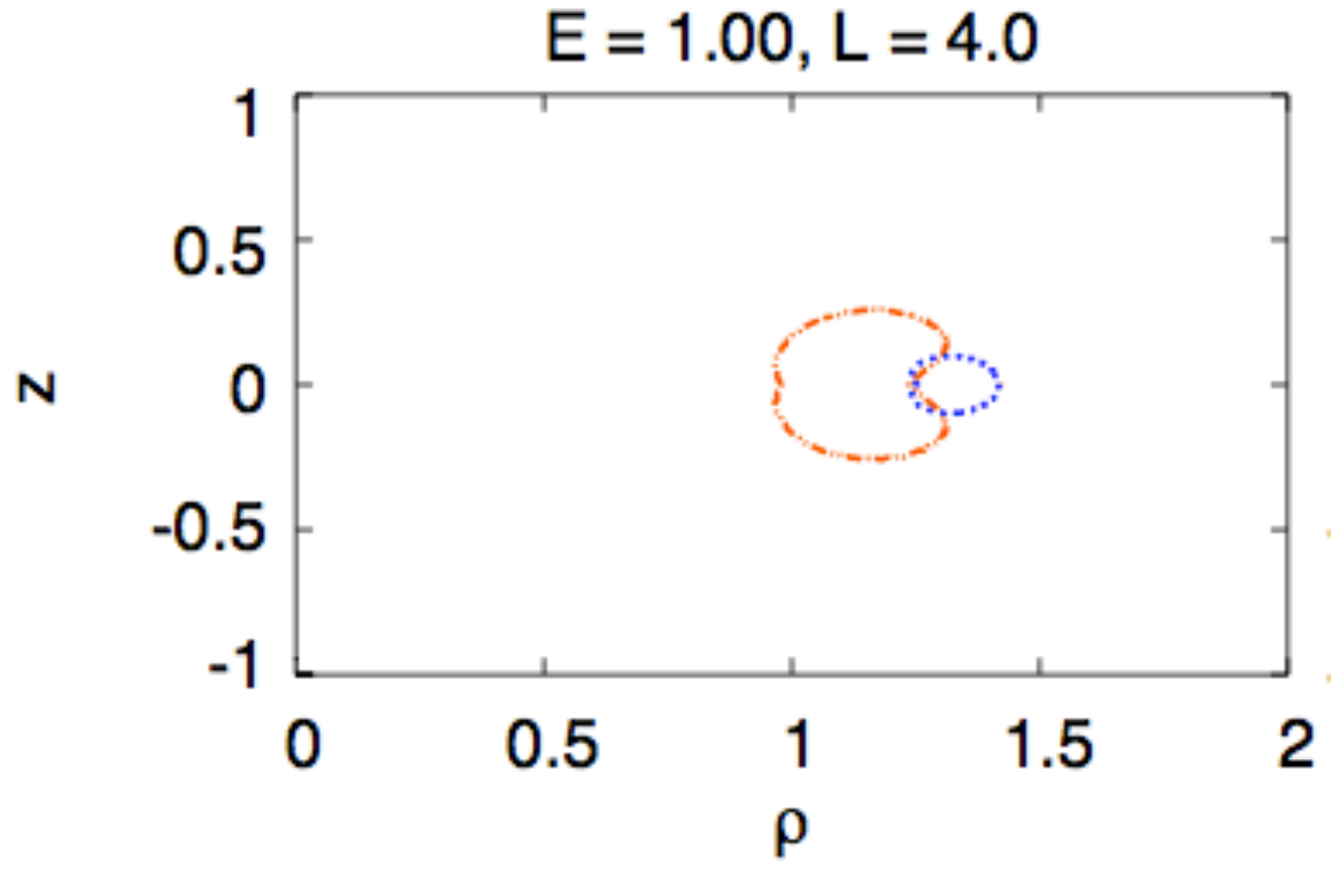}
\end{center}
\par
\vspace{-5mm} 
\caption{Effective potential for geodesic motion around a 
super-spinning object with $a_* = 1.2$ and $\tilde{q} = 1.0$
(solution MMS1) for $E=1.00$ and $L=3.1$ (left panels),
$L=3.6$ (central panels), and $L=4.0$ (right panels). Top panels: 
$xy$-plane. Bottom panels: enlargement of the region closer
to the object on the $\rho z$-plane. The dashed-dotted orange 
curves denote the zeros of the effective potential; dotted 
blue curves indicate the boundary of the causality violating 
region, where $g_{\phi\phi} < 0$. $L$, $\rho$, and $z$ are given 
in units of $M=1$.}
\label{f-3-2}
\end{figure}

\begin{figure}
\par
\begin{center}
\includegraphics[type=pdf,ext=.pdf,read=.pdf,width=5cm]{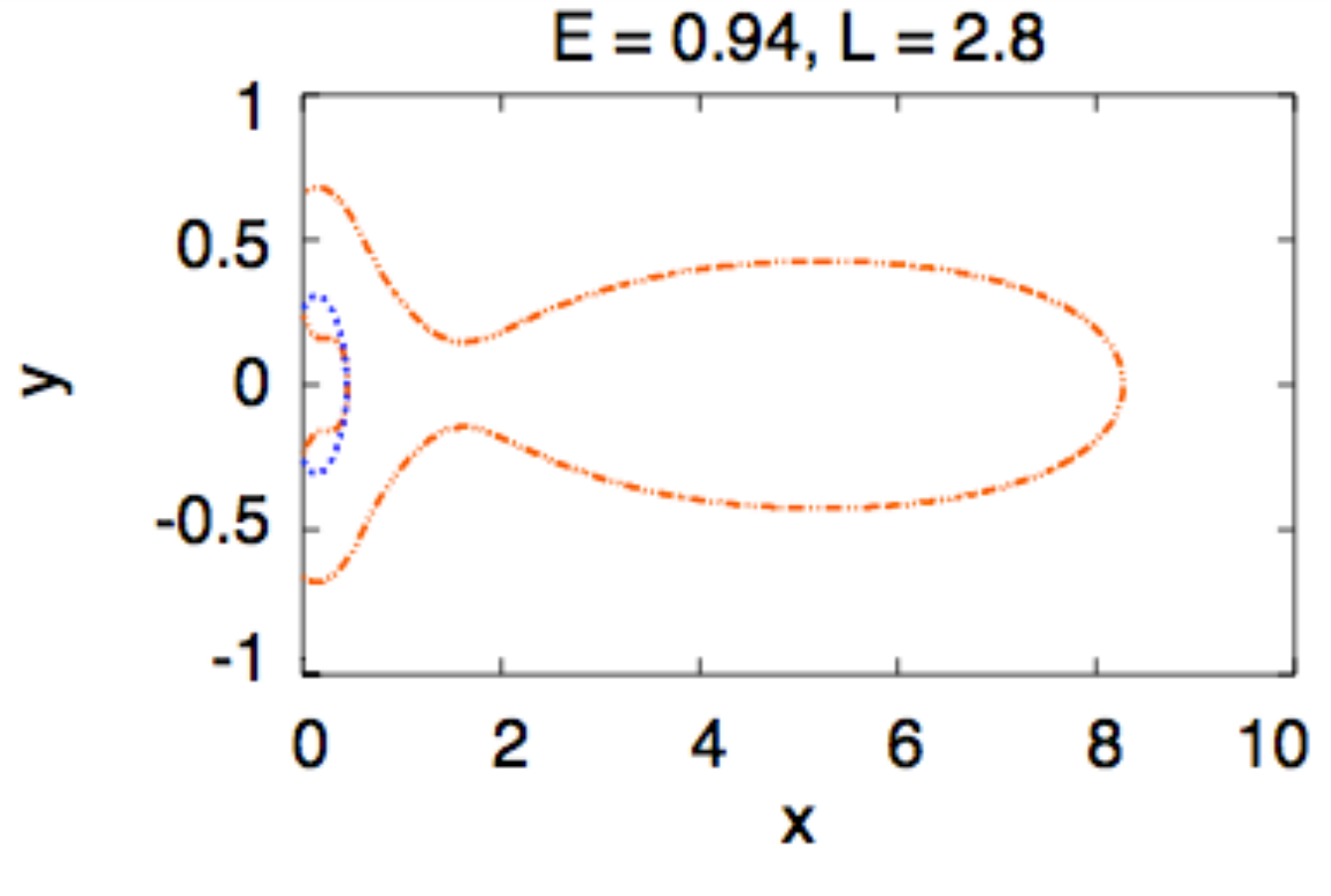}
\includegraphics[type=pdf,ext=.pdf,read=.pdf,width=5cm]{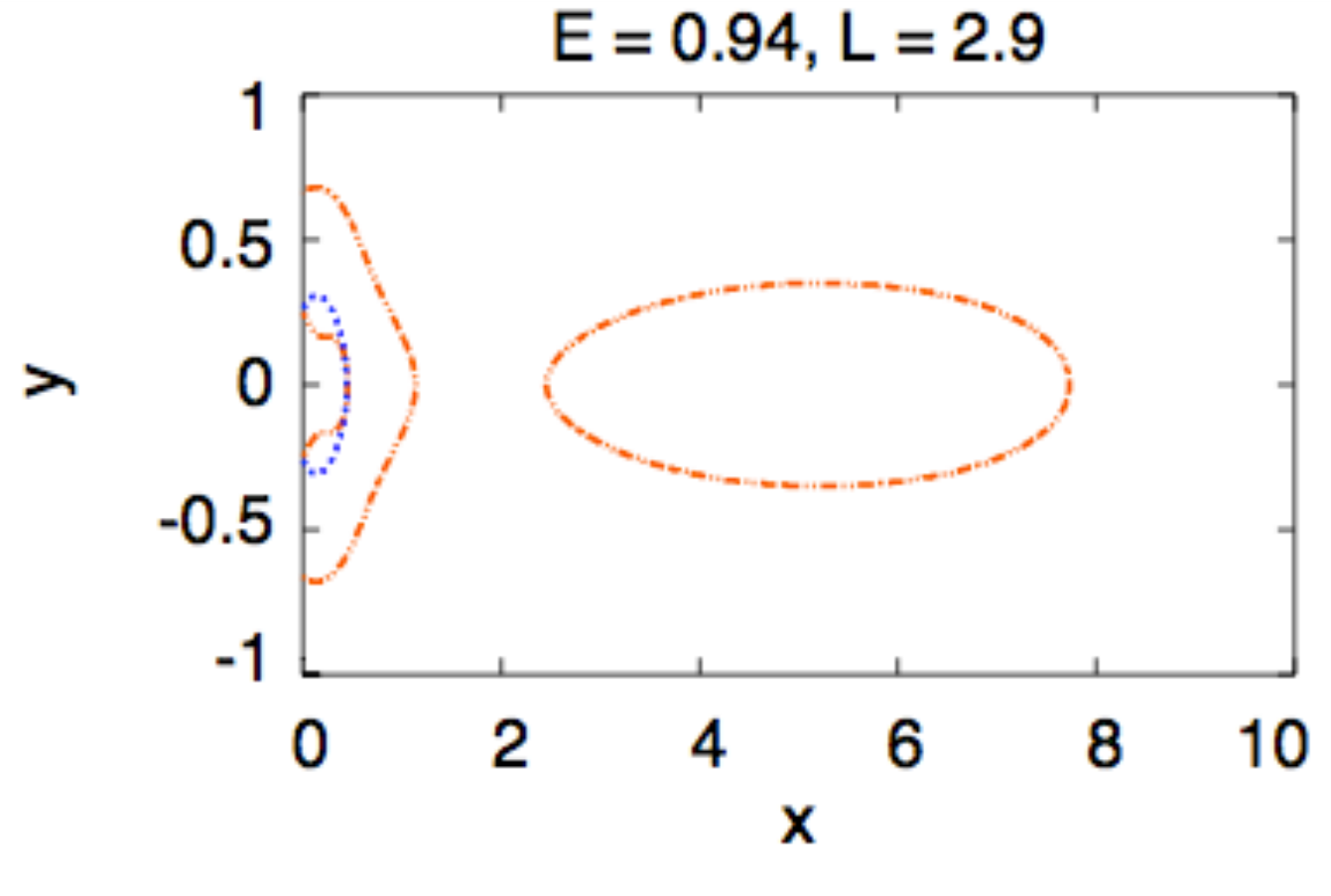}
\includegraphics[type=pdf,ext=.pdf,read=.pdf,width=5cm]{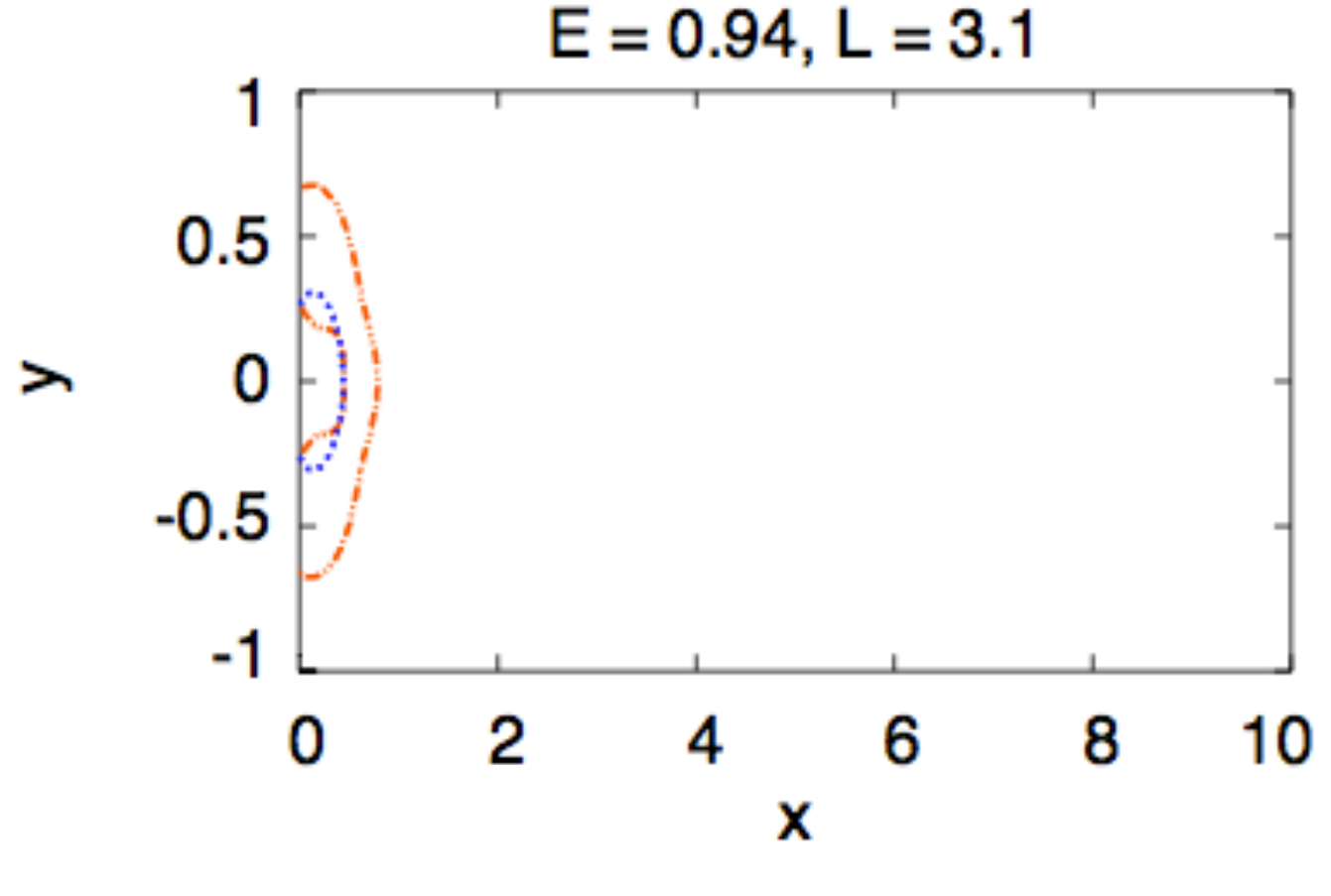}
\end{center}
\par
\vspace{-5mm} 
\caption{As in Fig.~\ref{f-3-2}, for $E=0.94$ and $L=2.8$
(left panel), $L=2.9$ (central panel), and $L=3.1$ (right panel)
on the $xy$-plane.}
\label{f-3-3}
\end{figure}

The MMS space-time is stationary and axisymmetric. There are 
thus two constants of motion associated respectively with the 
$t$- and $\phi$-coordinate; that is, the specific energy $E$ 
and the specific axial component of the angular momentum $L$:
\be\label{eq-e-l}
E = -g_{tt}\dot{t} - g_{t\phi}\dot{\phi} \, , \quad
L = g_{t\phi}\dot{t} + g_{\phi\phi}\dot{\phi} \, ,
\ee
where the dot denotes the derivative with respect an affine
parameter. From~(\ref{eq-e-l}), we have
\be
\dot{t}=\frac{E g_{\phi\phi} + L g_{t\phi}}{g_{t\phi}^2 
- g_{tt}g_{\phi\phi}} \, , \quad
\dot{\phi}=-\frac{E g_{t\phi} + Lg_{tt}}{g_{t\phi}^2 
- g_{tt}g_{\phi\phi}} \, .
\ee
By substituting $\dot{t}$ and $\dot{\phi}$ into the
equation of the conservation of the rest mass,
$g_{\mu\nu}\dot{x}^\mu\dot{x}^\nu = -1$, we find
\be\label{eq-veff}
\frac{e^{2\gamma}}{f}\left(\frac{\dot{x}^2}{x^2 + 1} + 
\frac{\dot{y}^2}{1 - y^2}\right) =
\frac{e^{2\gamma}}{f}(\dot{\rho}^2 + 
\dot{z}^2) = V_{\rm eff}(\rho,z,E,L) \, 
\ee
where 
\be
V_{\rm eff} = \frac{E^2}{f} -\frac{f}{\rho^2}
(L - \omega E)^2 - 1 \, .
\ee
Since the left-hand side of Eq.~(\ref{eq-veff}) is
non-negative, the motion of a test-particle is restricted
to the region $V_{\rm eff}\ge0$. The zeros of the
effective potential $V_{\rm eff}$ are shown in 
Figs.~\ref{f-3-2} and \ref{f-3-3} (orange dashed-dotted
curves) for $a_* = 1.2$, $\tilde{q}=1.0$ (MMS1), and 
different values of $E$ and $L$. They indicate the 
boundary of the allowed regions of the motion. The blue
dotted curves denote the boundaries of the closed time-like 
curve regions and must be inside the object. 
In Fig.~\ref{f-3-2}, top panels, oblate 
spheroidal coordinates $xy$ are used. The bottom panels 
are enlargements of the region close to the compact object 
in quasi-cylindrical coordinates $\rho z$.

For $E=1$ and large $L$ (right panels in Fig.~\ref{f-3-2}),
we find an unbound region extending to infinity and a
plunging region close to the massive object. As $L$
decreases, the unbound region moves closer to the
plunging region and a new allowed region of motion appears
(central panels of Fig.~\ref{f-3-2}). The latter has no
counterpart in the Kerr space-time, while it exists in the 
MN case~\cite{emri3}. For smaller $L$ (left panels of 
Fig.~\ref{f-3-2}), the unbound region merges with the 
plunging one.

For $E<1$, the situation is similar. 
In general, one find a bound region and a plunging
region (central panel of Fig.~\ref{f-3-3}). As $L$
decreases, the bound region approaches the plunging
region and eventually merges with the latter (left panel
of Fig.~\ref{f-3-3}). On the other hand, for larger
values of $L$ there is the plunging region only (right
panel of Fig.~\ref{f-3-3}).

In conclusion, the behavior of the geodesic motion is
not very different from the Kerr and MN space-times. In
particular, one can expect that the standard picture
of the accretion process is correct: the particles of the gas
approach the compact objects by losing energy and
angular momentum. When they are at the innermost stable
circular orbit (ISCO), they quickly plunge to the 
object\footnote{This picture is surely correct for $\tilde{q} 
\ge 0$, as the radius of the ISCO is always determined by 
the orbital stability along the radial direction. For 
$\tilde{q} < 0$, the radius of the ISCO may depend on the 
stability of the orbit along the vertical direction, see e.g. 
the discussion in Refs.~\cite{noi,ss}. In this second case, 
the gas particles reach the ISCO and then leave the equatorial 
plane. They plunge to the compact object after having lost 
additional energy and angular momentum. Such an additional 
loss of energy and angular momentum seems to be negligible in 
the description of the accretion process, but that should be 
checked in future studies by performing hydrodynamical and
magnetohydrodynamical simulations.}.

In addition to $E$, $L$, and the rest mass, in the Kerr
space-time there is a fourth constant of motion, the Carter
constant ${\cal C}$~\cite{carterc}. Its existence allows for the 
full separability of the equations of motion and it is a major 
source of research in modern day EMRI modeling. For a generic
space-time, the existence of an effective fourth constant 
of motion can be checked by studying the Poincar\'e maps for 
geodesic motion of bound orbits: if they are all closed 
curves, there is a fourth constant of motion~\cite{emri3}.
In analogy with the case of the MN solution discussed 
in~\cite{emri3}, presumably even in the MMS space-time all 
orbits are strictly speaking chaotic, and no true fourth integral 
exists, but most of the orbits appear to be regular and 
one can find a quantity that is nearly invariant along them.

\section{Evolution of the spin parameter \label{s-4}}

In this section, I compute the evolution of the spin parameter
of the compact object due to the accretion process from a thin 
disk following Ref.~\cite{bardeen}. I assume that the disk is 
on the equatorial plane of the object\footnote{This assumption
is correct for long term accretion onto a super-massive object 
at the center of a galaxy, since the alignment timescale of 
the spin of the object with the disk is typically much shorter 
than the accretion timescale~\cite{ss}.} and that the disk's 
gas moves on nearly geodesic circular orbits. The gas falls to 
the central object by loosing energy and angular momentum. 
When it reaches the ISCO, it 
plunges to the massive body. If the gas is ``absorbed'' by the 
compact object, with no further emission of radiation, the 
compact object changes its mass by $\delta M = E_{\rm ISCO} 
\delta m$ and its spin angular momentum by $\delta J = L_{\rm 
ISCO} \delta m$, where $E_{\rm ISCO}$ and $L_{\rm ISCO}$ are 
respectively the specific energy and the specific angular 
momentum of the gas particle at the ISCO, while $\delta m$ 
is the gas rest-mass. The evolution of the spin parameter 
turns out to be governed by the following equation~\cite{bardeen}
\be\label{eq-a}
\frac{da_*}{d\ln M} = \frac{1}{M} 
\frac{L_{\rm ISCO}}{E_{\rm ISCO}} - 2 a_* \, .
\ee

In the case of a body with a solid surface made of ordinary 
matter (i.e. protons, neutrons, and electrons), this picture is 
not rigorously correct: the gas particles hit the surface of 
the body and release their gravitational energy in form of 
radiation. They are also accumulated on the surface and may
undergo nuclear reactions, with the production of bursts. 
In such a situation, the computation of the evolution of the
spin parameter is much more complicated and model dependent.
On the contrary, if the compact
object is a BH, the gas particles are really absorbed without
further emission of radiation, since the BH has no solid 
surface, and, once the particles are behind the event horizon, 
their radiation cannot escape to infinity. For a BH, one can 
analytically integrate Eq.~(\ref{eq-a}) and obtain~\cite{bardeen}
\be
a_* &=&  \sqrt{\frac{2}{3}}
\frac{M_0}{M} \left[4 - \sqrt{18\frac{M_0^2}{M^2} - 2}\right] 
 \quad {\rm for} \, M \le \sqrt{6} M_0 \, , \nonumber\\
a_* &=&  1  \quad {\rm for} \, M > \sqrt{6} M_0 \, ,
\ee
assuming an initially non-rotating BH with mass $M_0$.
The equilibrium is reached for $a_*^{eq} = 1$ after the BH has
increased its mass by a factor $\sqrt{6} \approx 2.4$. Including the 
effect of the radiation emitted by the disk and captured by the 
BH, one finds $a_*^{eq} \approx 0.998$~\cite{thorne}, because 
radiation with angular momentum opposite to the BH spin has 
larger capture cross section. The presence of magnetic fields in 
the plunging region may further reduce this value to $a_*^{eq} 
\sim 0.95$~\cite{gammie,sh05}, by transporting angular momentum 
outward.

For astrophysical BH candidates, we can presumably use 
Eq.~(\ref{eq-a}). They indeed seems to be capable of absorbing 
the accreting gas without apparent loss of energy and angular 
momentum: no electromagnetic emission from their surface is 
observed, see e.g. Refs.~\cite{eh0,eh1,eh2}. For a generic
stationary and axisymmetric space-time, $E_{\rm ISCO}$ and 
$L_{\rm ISCO}$ can be computed numerically, as described 
in~\cite{noi}. Here I assume that $\tilde{q}$ is a constant
and does not change as the object increases its mass and
is spun up, but this is not true in general. For example, in 
the case of neutron stars, $\tilde{q}$ depends on the mass of 
the body~\cite{neutron}. If the astrophysical BH candidates
were objects with an anomalous quadrupole moment depending
strongly on the mass of the body, objects with different mass
could have a different equilibrium spin parameter.

\begin{figure}
\par
\begin{center}
\includegraphics[type=pdf,ext=.pdf,read=.pdf,width=7.5cm]{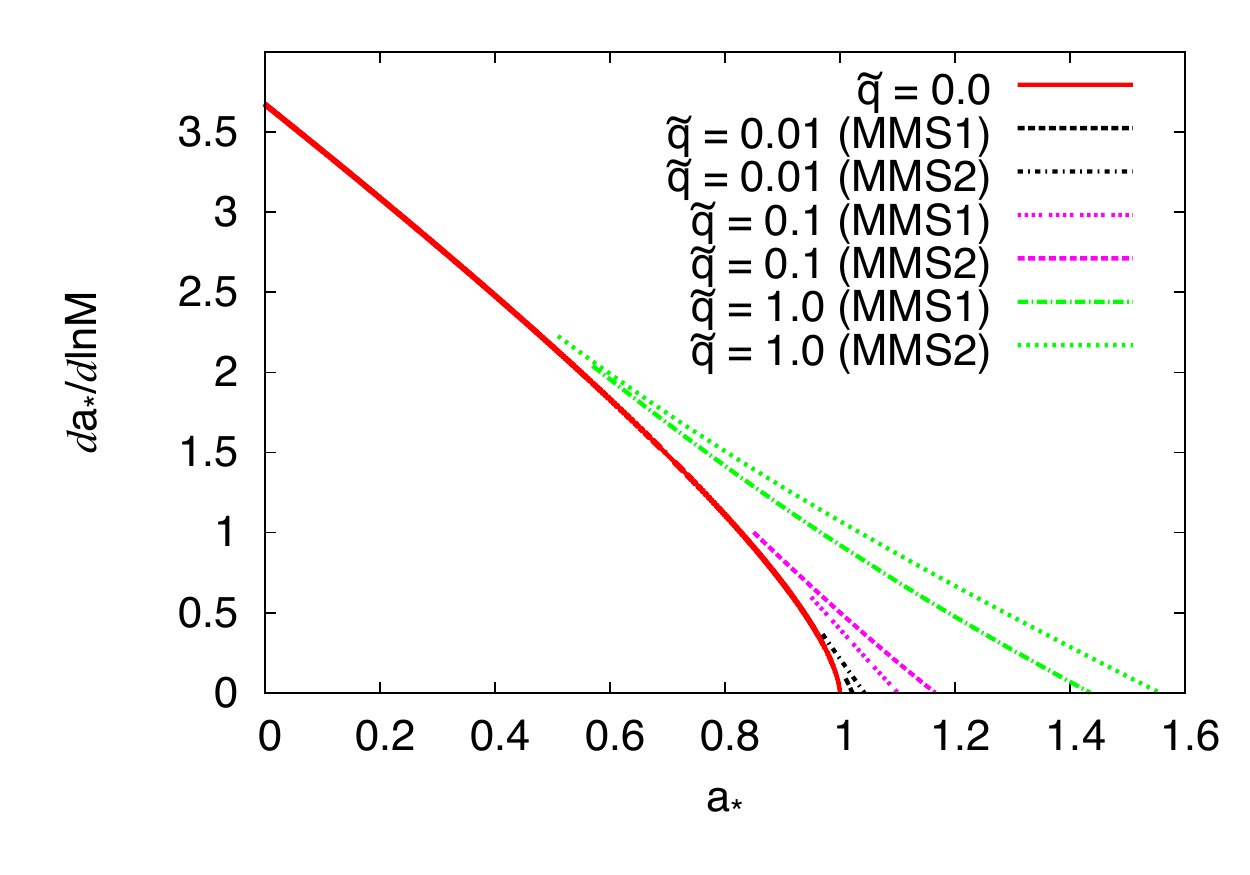}
\includegraphics[type=pdf,ext=.pdf,read=.pdf,width=7.5cm]{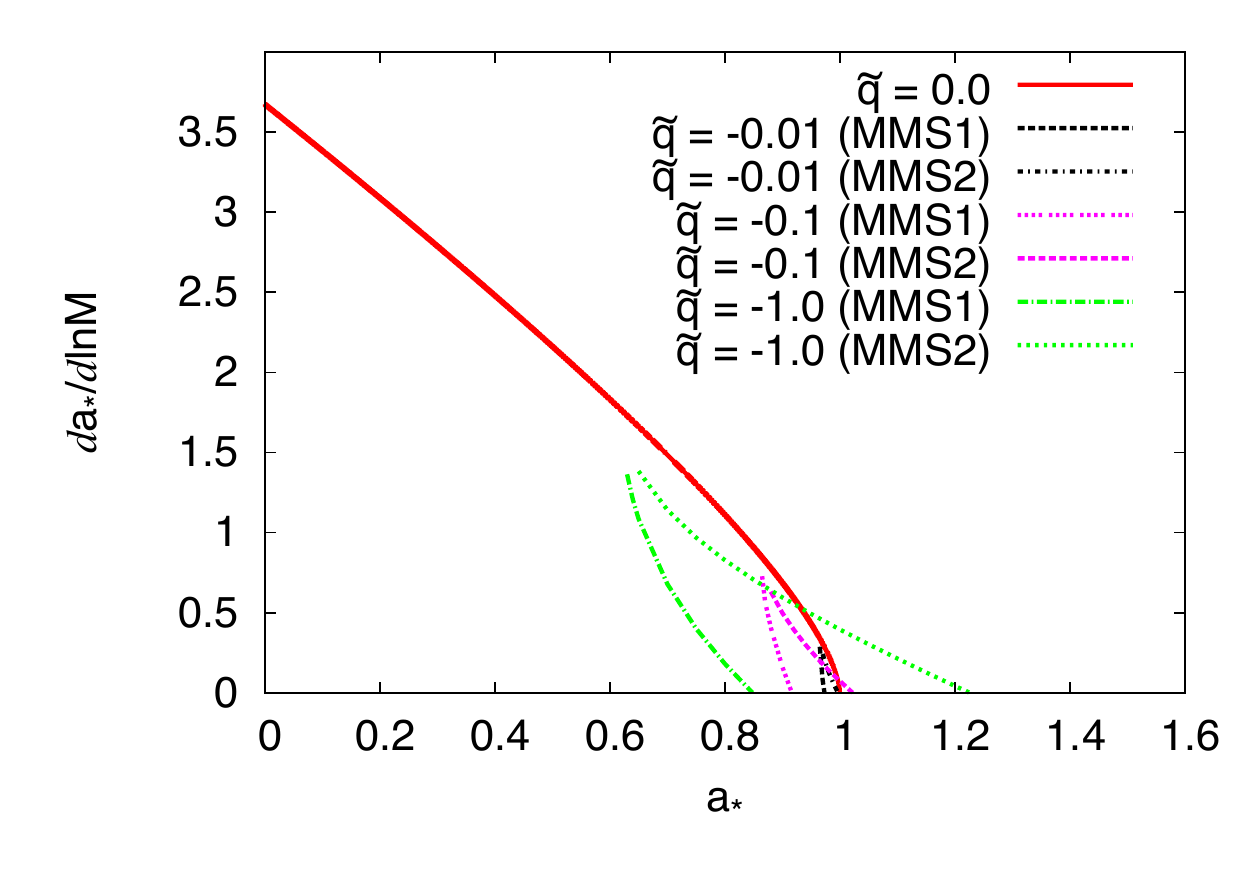}
\end{center}
\par
\vspace{-5mm} 
\caption{$da_*/d\ln M$ as a function of $a_*$ for
different value of the anomalous quadrupole moment $\tilde{q}$.
The red solid curve is for the case of a BH ($\tilde{q}=0$).
Left panel: cases with $\tilde{q}\ge 0$.
Right panel: cases with $\tilde{q}\le 0$.}
\label{f-4-1}
\end{figure}

\begin{figure}
\par
\begin{center}
\includegraphics[type=pdf,ext=.pdf,read=.pdf,width=7.5cm]{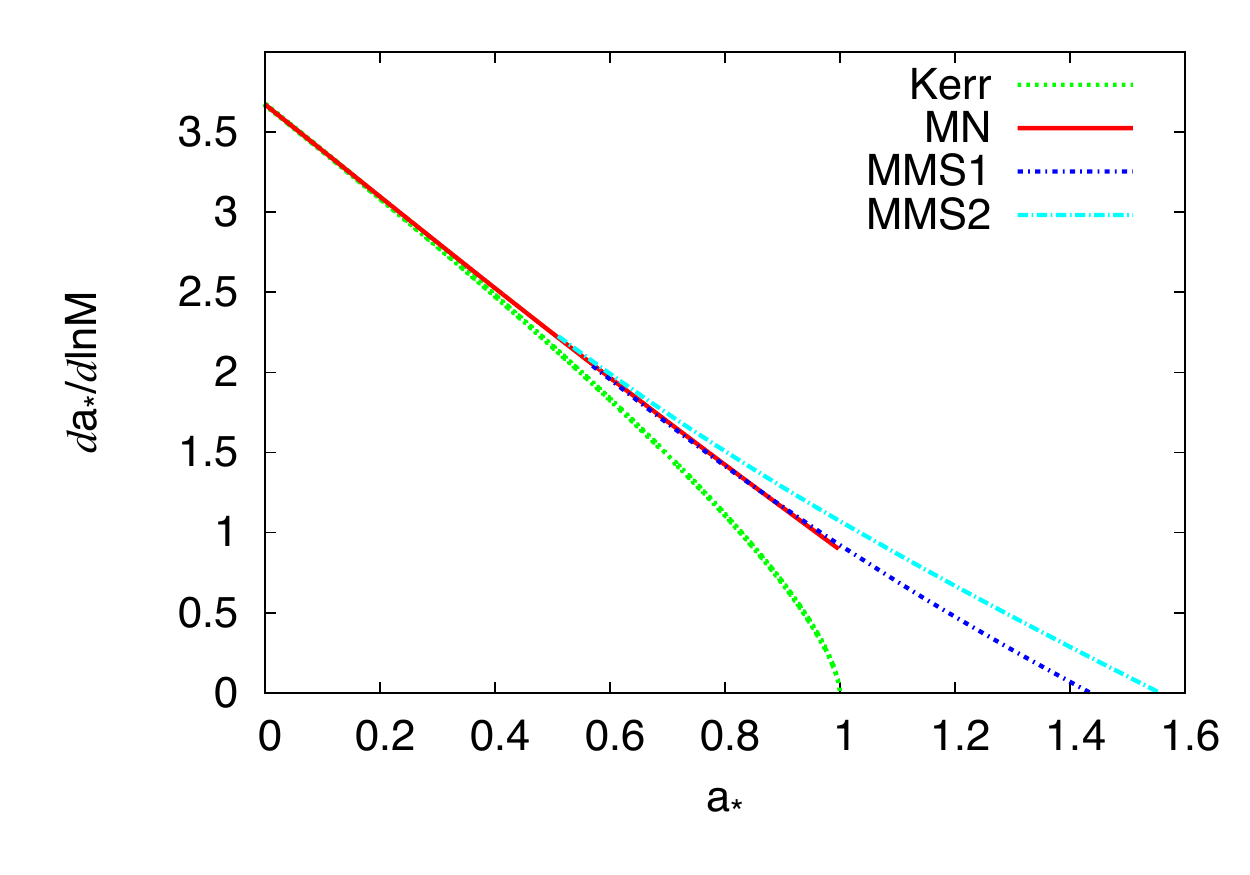}
\includegraphics[type=pdf,ext=.pdf,read=.pdf,width=7.5cm]{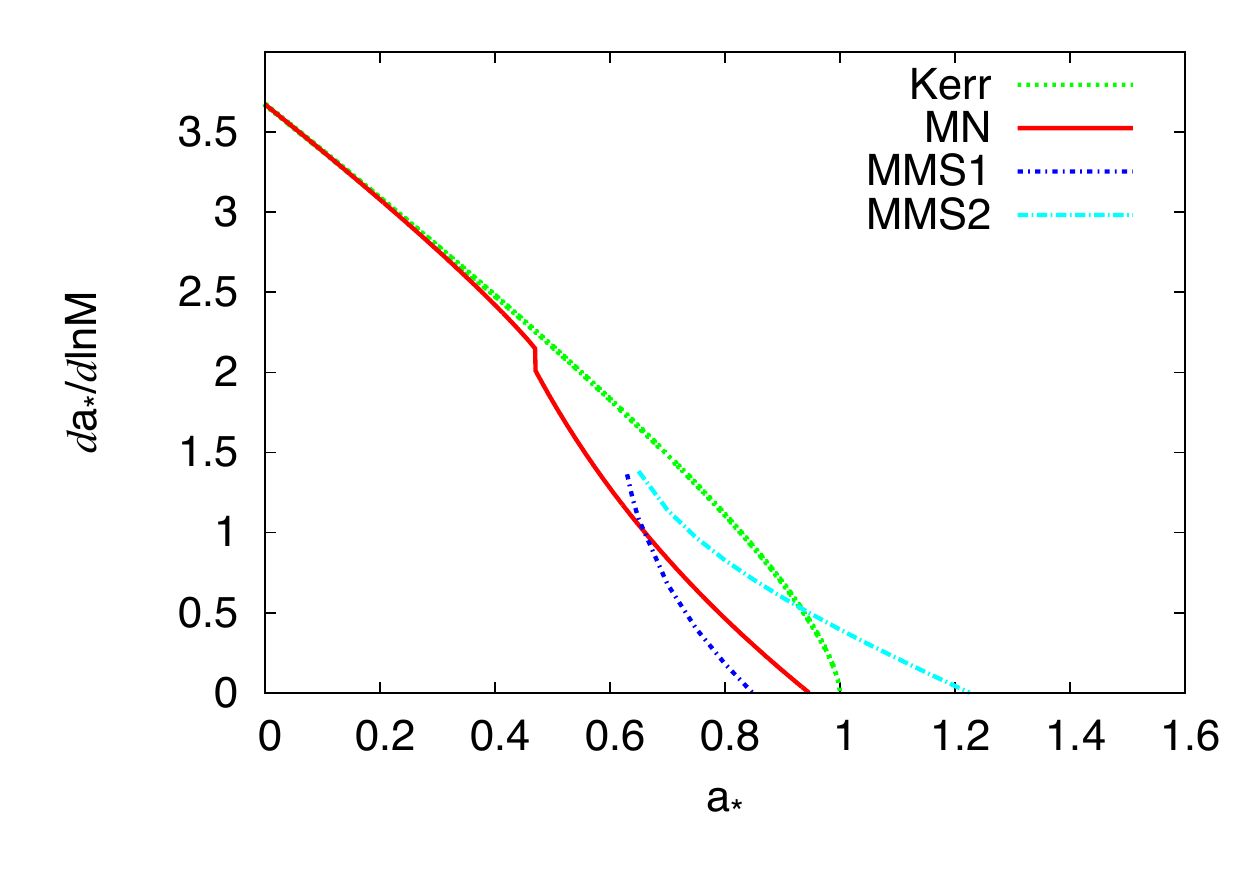}
\end{center}
\par
\vspace{-5mm} 
\caption{Comparison of the function $da_*/d\ln M$ for the 
case of the MN solution with the MMS1 and MMS2 solution with the 
same value of the anomalous quadrupole moment. Left panel: 
$\tilde{q} = 1$. Right panel: $\tilde{q} = -1$. The MN solution 
can be used till $a_* = 1$. The green dotted curve is for the 
case of a BH.}
\label{f-4-2}
\end{figure}

Neglecting the effect of the radiation captured by the object
and the presence of magnetic fields in the plunging region,
in Fig.~\ref{f-4-1} I show $da_*/d\ln M$ as a function of 
the spin parameter $a_*$ for some values of the anomalous 
quadrupole moment $\tilde{q}$. The value of the equilibrium 
spin parameter $a_*^{eq}$ for these cases is reported in 
Tab.~\ref{t-eq}. In Fig.~\ref{f-4-1}, the solid red curve 
is for a BH ($\tilde{q}=0$) and has been computed by using 
Boyer--Lindquist coordinates. Indeed, even if the Kerr metric 
is included in the MMS solution, neither the form of the 
metric presented in~\cite{mms1,mms2}, nor the one discussed 
in Sec.~\ref{s-2} are suitable for numerical calculations, 
as they would require that the parameter $b$ is an imaginary 
number. For a similar reason, for a given $\tilde{q} \neq 0$ 
we cannot study the evolution of the spin parameter from 
$a_*=0$ to $a_*=a_*^{eq}$, but from some value $a_*^{in}>0$,
depending on $\tilde{q}$. When $da_*/d\ln M>0$, the accretion 
process spins the compact body up. When $da_*/d\ln M<0$, the 
compact body is spun down. The equilibrium spin parameter 
is thus the one for which $da_*/d\ln M=0$. In Fig.~\ref{f-4-2},
I show the quantity $da_*/d\ln M$ for a BH (green dotted curve)
and for three compact objects with the same anomalous quadrupole
moment $\tilde{q}=\pm1.0$, but different higher order moments:
the subclass of MN metrics discussed in Ref.~\cite{ss} (red 
solid curve), MMS1 (dark-blue dotted-dashed curve), and MMS2 
(light-blue dotted-dashed curve).

\begin{figure}
\par
\begin{center}
\includegraphics[type=pdf,ext=.pdf,read=.pdf,width=7.5cm]{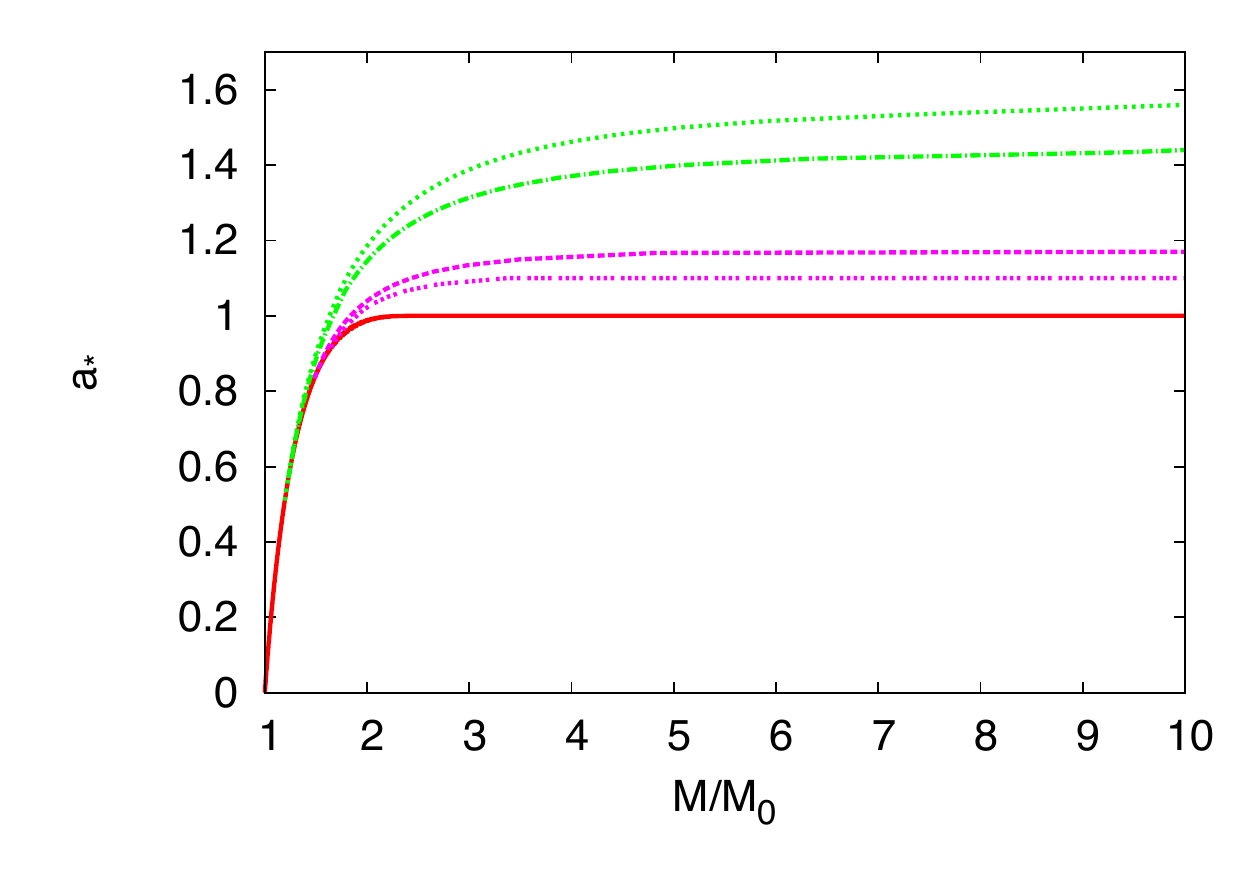}
\includegraphics[type=pdf,ext=.pdf,read=.pdf,width=7.5cm]{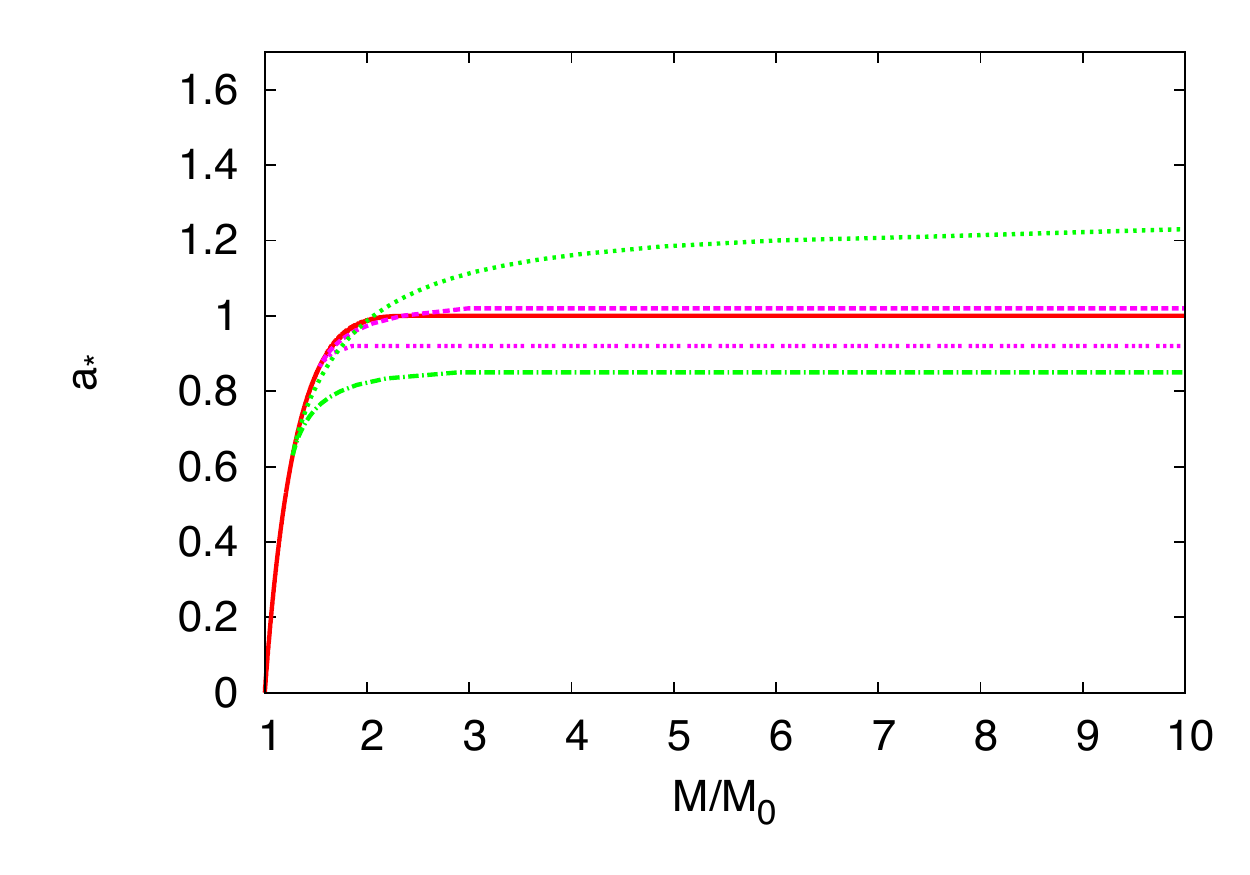}
\end{center}
\par
\vspace{-5mm} 
\caption{Evolution of the spin parameter $a_*$ as a function of 
$M/M_0$. The color and the style of the curves for any $\tilde{q}$ 
is the same of Fig.~\ref{f-4-1} (the cases $\tilde{q} = \pm 0.001$
are not shown because for them the evolution of $a_*$ can be 
studied only in a very limited interval of the spin parameter).
The red solid curve is for the case of a BH ($\tilde{q}=0$).
Left panel: cases with $\tilde{q}\ge 0$.
Right panel: cases with $\tilde{q}\le 0$.}
\label{f-4-3}
\end{figure}

Fig.~\ref{f-4-3} shows the evolution of the spin parameter
$a_*$ as a function of $M/M_0$, the mass to initial mass ratio.
Eq.~(\ref{eq-a}) is numerically integrated from $a_*^{in}$, 
with $M/M_0$ equal to the one of the Kerr solution with the
same spin parameter.

\begin{table}
\begin{center}
\begin{tabular}{c c c c c c c}
\hline \\
$\tilde{q}$ & \hspace{.5cm} & $a_*^{eq}$ (MMS1) & \hspace{.5cm} & 
$a_*^{eq}$ (MMS2) & \hspace{.5cm} & $a_*^{eq}$ (Kerr) \\ \\
\hline 
1.0 & & 1.44 & & 1.56 & & -- \\
0.1 & & 1.10 & & 1.17 & & -- \\ 
0.01 & & 1.02 & & 1.04 & & -- \\ 
0.0 & & -- & & -- & & 1.00 \\ 
-0.01 & & 0.97 & & 0.997 & & -- \\ 
-0.1 & & 0.92 & & 1.02 & & -- \\ 
-1.0 & & 0.85 & & 1.23 & & -- \\
\hline
\end{tabular}
\end{center}
\caption{Equilibrium spin parameter $a_*^{eq}$ for the cases
shown in Fig.~\ref{f-4-1}.}
\label{t-eq}
\end{table}

\section{Discussion \label{s-5}}

\begin{figure}
\par
\begin{center}
\includegraphics[type=pdf,ext=.pdf,read=.pdf,width=7.5cm]{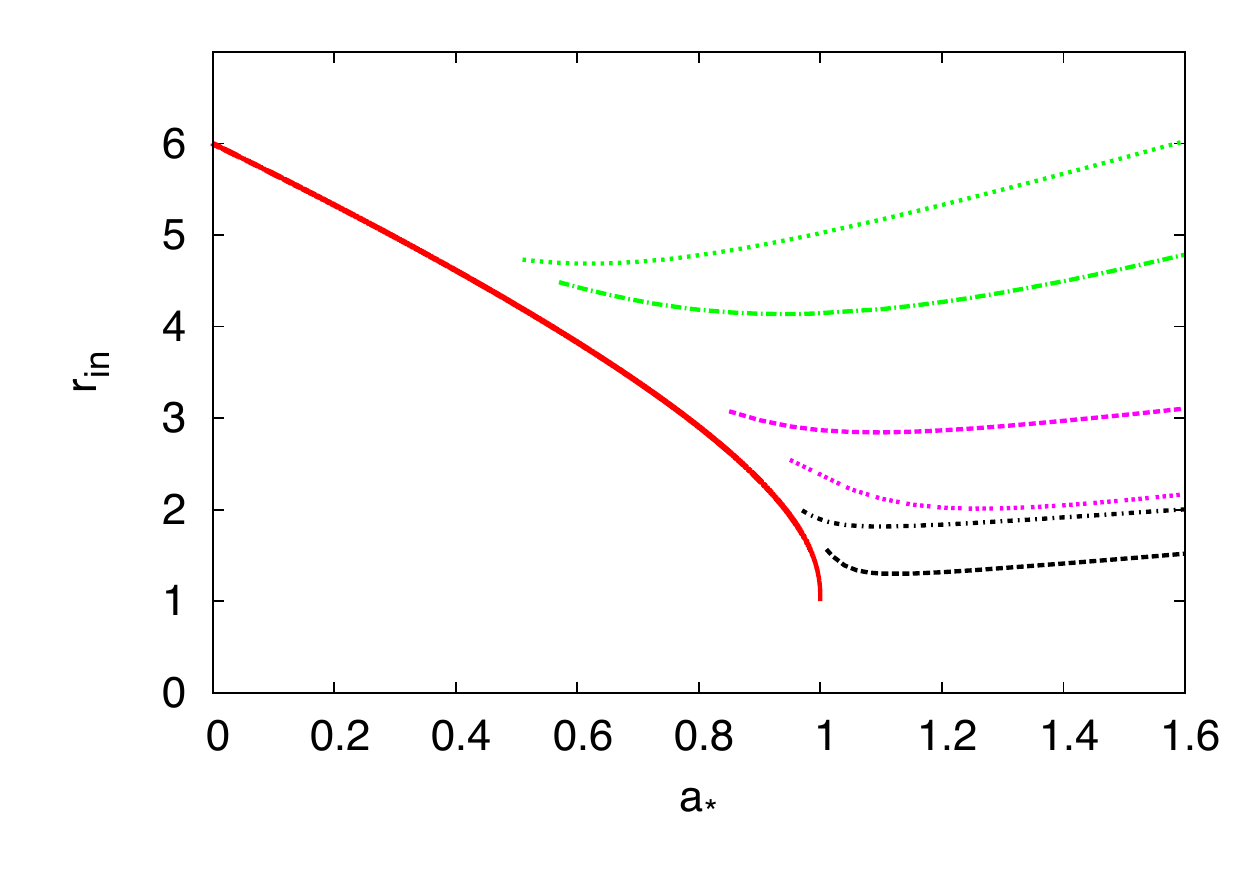}
\includegraphics[type=pdf,ext=.pdf,read=.pdf,width=7.5cm]{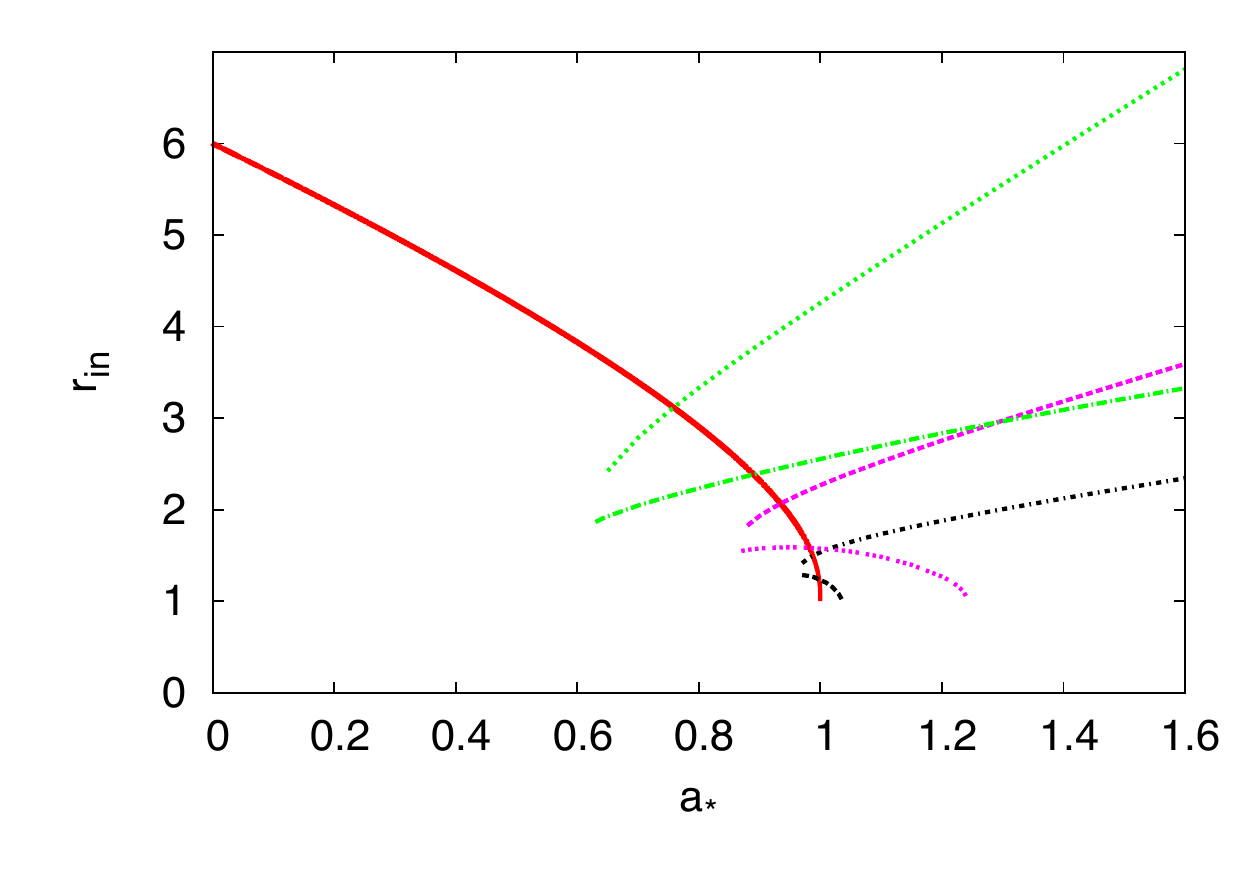}
\end{center}
\par
\vspace{-5mm} 
\caption{Inner radius of the accretion disk as a 
function of the spin parameter $a_*$ for different values of 
the anomalous quadrupole moment $\tilde{q}$. The color and the 
style of the curves for any $\tilde{q}$ is the same of
Fig.~\ref{f-4-1}. Left panel: cases with $\tilde{q}\ge0$. 
Right panel: cases with $\tilde{q}\le0$. Inner radius of the 
disk in Schwarzschild coordinates and in units of $M=1$.}
\label{f-5-1}
\par
\begin{center}
\includegraphics[type=pdf,ext=.pdf,read=.pdf,width=7.5cm]{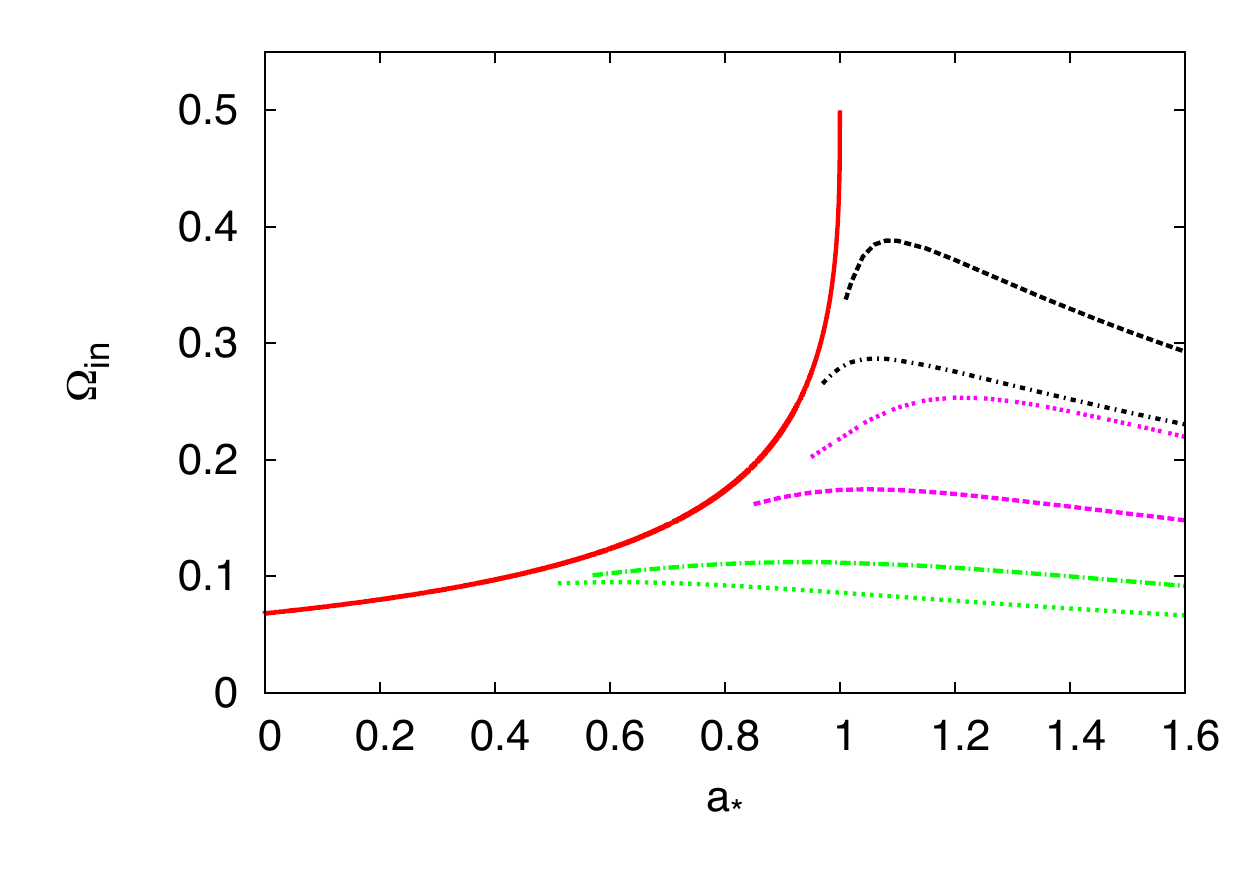}
\includegraphics[type=pdf,ext=.pdf,read=.pdf,width=7.5cm]{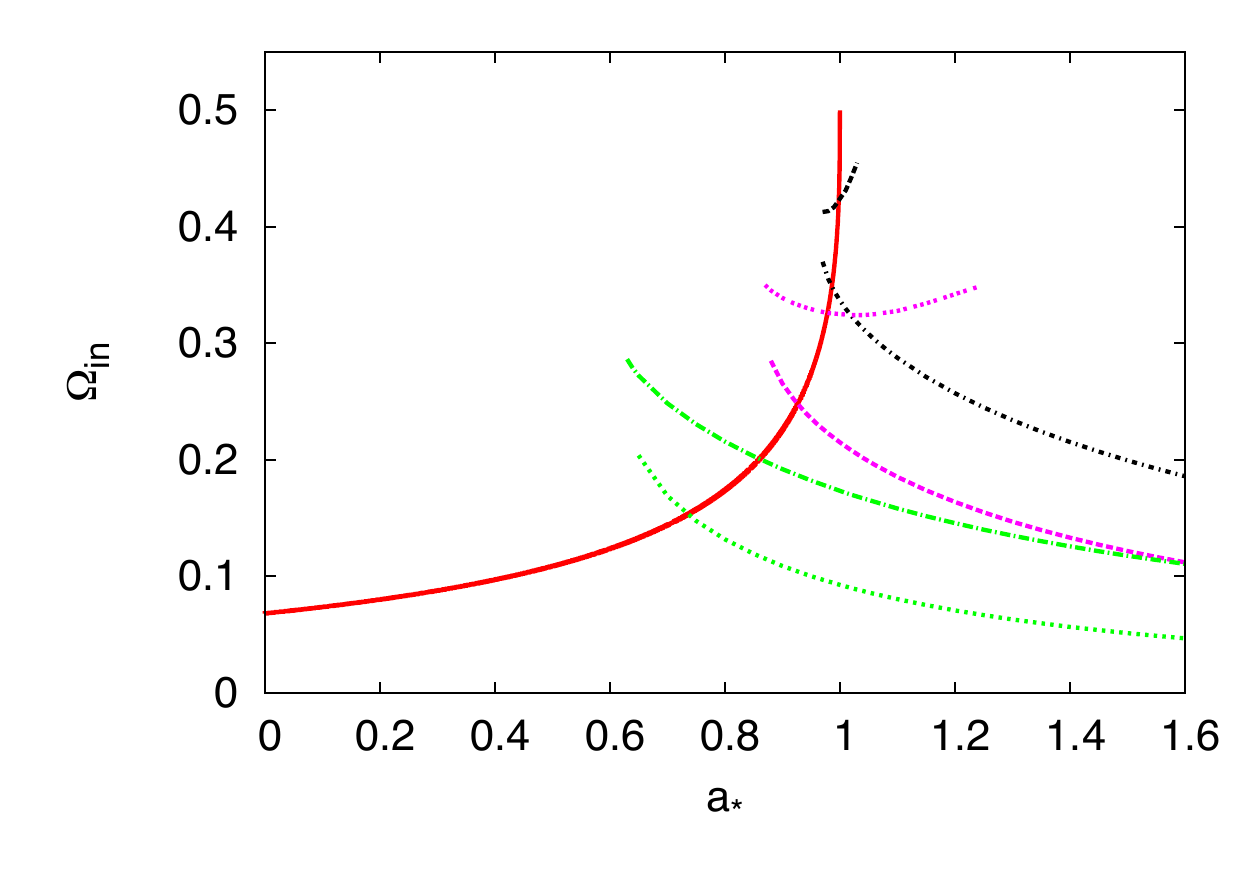}
\end{center}
\par
\vspace{-5mm} 
\caption{Angular velocity at the inner radius of the accretion 
disk as a function of the spin parameter $a_*$ for different values 
of the anomalous quadrupole moment $\tilde{q}$. The color and the 
style of the curves for any $\tilde{q}$ is the same of
Fig.~\ref{f-4-1}. Left panel: cases with $\tilde{q}\ge0$. 
Right panel: cases with $\tilde{q}\le0$. $\Omega_{in}$ is given
in units of $M=1$.}
\label{f-5-2}
\end{figure}

As the values of $E_{\rm ISCO}$ and $L_{\rm ISCO}$ in Eq.~(\ref{eq-a})
depend on the metric of the space-time, compact objects with 
different quadrupole (or higher) moment have a different 
equilibrium spin parameter $a_*^{eq}$. Figs.~\ref{f-4-1}, \ref{f-4-3}, 
and Tab.~\ref{t-eq} can be qualitatively understood in term of the 
inner radius of the disk $r_{in}$: for a given spin parameter,
$da_*/d\ln M$ depends on $L_{\rm ISCO} / E_{\rm ISCO}$, which typically
increases/decreases if $r_{in}$ is larger/smaller. Fig.~\ref{f-5-1}
shows the value of the inner radius of the disk in Schwarzschild
coordinates (see Eq.~(\ref{eq-rad}) in App.~\ref{a-1}) as a 
function of the spin parameter. For instance, the MMS1 solutions 
with $\tilde{q}<0$ have $a_*^{eq} < 1$; as we can see, their 
inner disk's radius is smaller than the one in Kerr space-time 
for $|a_*| \le a_*^{eq}$. The curves for the MMS1 solutions 
$\tilde{q} = -0.01$ and $-0.1$ in the right panel of
Fig.~\ref{f-5-1} stop at $r_{in}=M$, which is equivalent to
$x=0$, because our coordinates cannot describe the space-time 
at smaller radii ($0\le x < +\infty$). These two cases have
anyway to be taken with caution: there are regions with closed 
time-like curves with Schwarzschild radius larger than $M$, 
so even the surface of the compact object should be probably larger 
than $M$. Let us notice, however, that this happens for 
$a_*>a_*^{eq}$ and it can be neglected in the study of the 
evolution of the spin. As shown in Fig.~\ref{f-5-1}, for 
sufficiently high values of the spin parameter, the inner radius 
of the disk always increases as $a_*$ increases. This is not a 
surprise, because even in the Kerr space-time the radius of the 
ISCO reaches a minimum for $a_* \approx 1.089$, and then it 
increases as $a_*$ increases~\cite{harada}.

The value of the inner radius of the disk depends inevitably
on the choice of the coordinates; the comparison of this quantity
for objects with different value of $\tilde{q}$ is thus not 
so meaningful. From this point of view, a more interesting
quantity is the angular velocity of a gas particle at the inner 
radius of the disk, which is shown in Fig.~\ref{f-5-2}. As we
can see, the angular frequency at $r_{in}$ can be very high 
around a very fast-rotating BH, while it is significantly lower 
for another compact object, regardless of its spin parameter. 
This can be an interesting observational feature to test the 
Kerr nature of astrophysical BH candidates, because it implies
that it is very difficult for an object that is not a BH to
mimic a very fast-rotating BH. In other words, among the
stationary and axisymmetric space-times, the Kerr solution is 
a very special case with peculiar properties and this is
particularly true for $a_* \rar 1$. Generally speaking, the
accretion process onto an object with non-Kerr quadrupole moment
looks more like the accretion onto a slow-rotating BH, but it can
hardly present some features of a fast-rotating BH.

\begin{figure}
\par
\begin{center}
\includegraphics[type=pdf,ext=.pdf,read=.pdf,width=7.5cm]{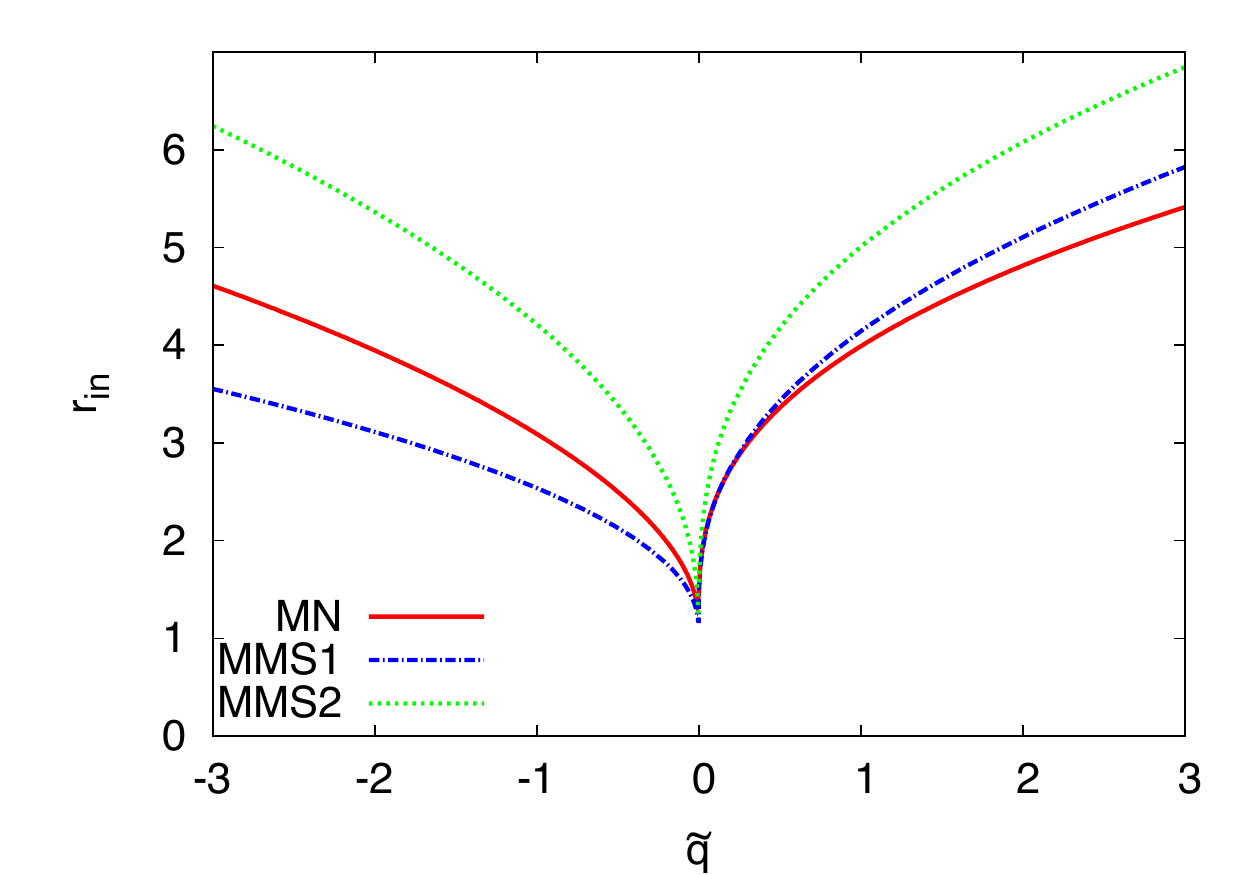}
\includegraphics[type=pdf,ext=.pdf,read=.pdf,width=7.5cm]{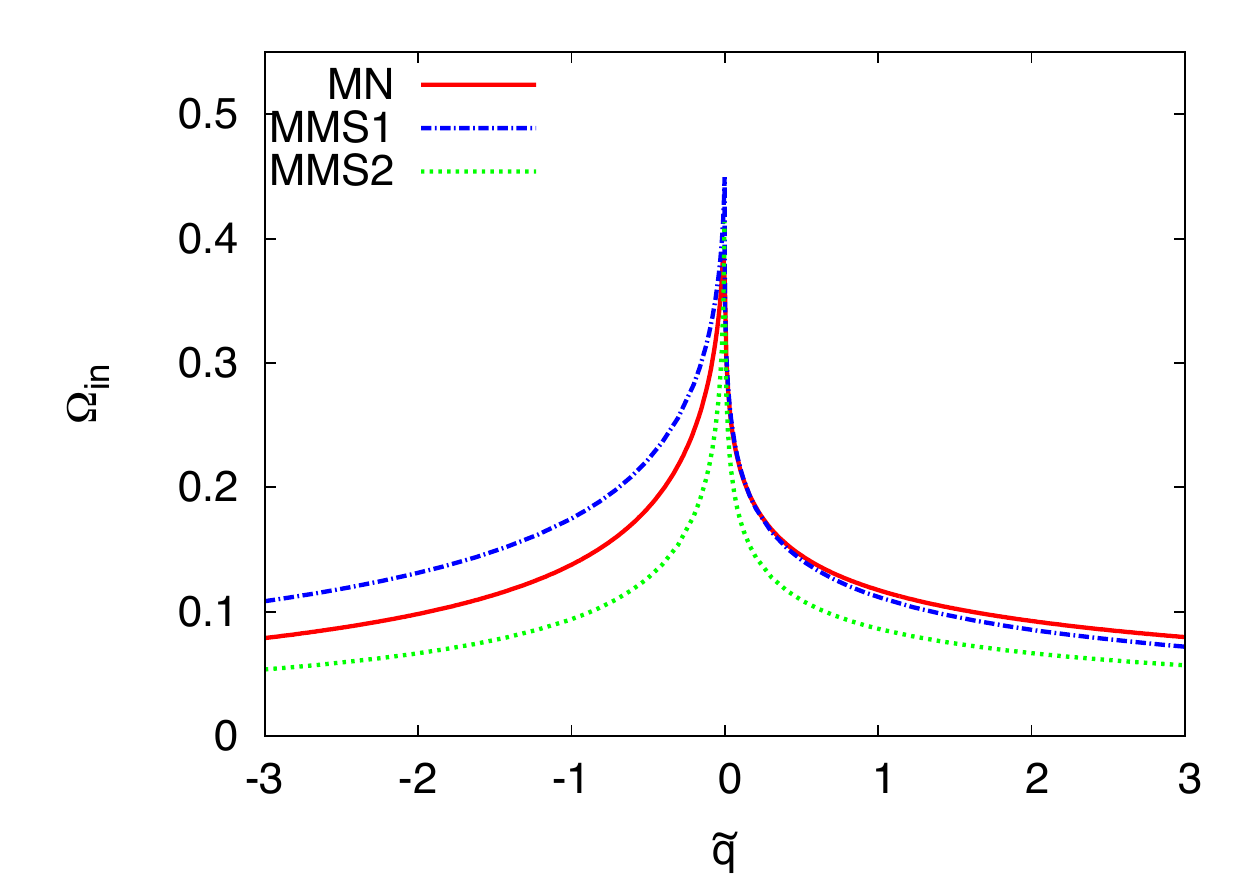}
\end{center}
\par
\vspace{-5mm} 
\caption{Left panel: inner radius of the disk for bodies with
spin parameter $a_* = 0.99$ as a function of the anomalous 
quadrupole moment $\tilde{q}$ for the MN (red solid curve), 
MMS1 (blue dashed-dotted curve), and MMS2 (green dotted curve) 
solutions. Right panel: as in the left panel, for the angular
velocity at the inner radius of the disk. Inner radius of the 
disk in Schwarzschild coordinates and in units of $M=1$. 
$\Omega_{in}$ is given in units of $M=1$.}
\label{f-5-3}
\end{figure}

For fast-rotating objects, even deformations beyond the 
quadrupole moments are important for the properties of the 
space-time. In Fig.~\ref{f-5-3}, I show the inner radius of
the disk and the corresponding angular frequency of a gas
particle around objects with $a_*=0.99$ as a function of the
anomalous quadrupole moment. The three curves represent three
objects with same mass, spin, and quadrupole moment, but
different higher order moments: the subclass of MN solutions 
discussed in~\cite{ss}, the MMS1 solution, and the MMS2 solution.

As already noticed in~\cite{ss}, Eq.~(\ref{eq-a}) provides the 
correct evolution of the spin parameter if the compact object
does not become unstable before reaching the equilibrium value
$a_*^{eq}$. More in general, the accretion process can spin the
body up, but there may exist other processes that spin it down.
This is what should happen, for instance, in the case of a 
neutron star: the gas of the accretion disk can spin the neutron
star up, but, when the rotational frequency of the object exceeds
$\sim 1$~kHz, there are unstable modes that spin the neutron
star down through the emission of gravitational waves~\cite{kkk}. 
If something similar happens for BH candidates, the maximum 
value of the spin parameter would be determined by the internal
structure of these objects. The latter may spin down by 
emitting gravitational waves, presumably as a burst, potentially 
detectable by future experiments.

The fact that the accretion process can spin a compact object up 
to $a_*>1$ can be relevant for the super-massive BH candidates 
at the center of galaxies, while it should be negligible for 
stellar-mass objects in X-ray binary systems. In general, the 
value of the spin parameter of a compact object is determined by 
the competition of three physical processes: the event creating 
the object, mergers, and gas accretion. For the stellar-mass BH 
candidates in X-ray binary systems, the value of their spin should 
reflect the one at the time of their creation. If they belong to 
low-mass X-ray binary systems, even swallowing the whole stellar 
companion they cannot change significantly their spin, because 
the mass of the stellar companion is much smaller. If they are 
in high-mass X-ray binary systems, even accreting at the Eddington 
limit they do not have enough time to grow before the explosion 
of the companion. On the contrary, for the super-massive objects 
in galactic nuclei the initial spin value is completely unimportant, 
as they have increased their mass by a few orders of magnitude 
from the original one. In the case of prolonged disk accretion, 
the object has the time to align itself with the disk and the 
process of gas accretion should dominate over mergers~\cite{m2}. 
This picture is supported even by current estimates of the 
mean radiative efficiency of AGN~\cite{ho}.
As shown in~\cite{ss}, the alignment timescale is the same 
for BHs and objects with non-Kerr quadrupole moment, as long 
as $|\tilde{q}| \ll 100$.

\section{Conclusions \label{s-6}}

The final product of the gravitational collapse of matter
is thought to be a black hole and there are astrophysical 
evidences for the existence of dark objects that are too
compact and too heavy to be relativistic stars or clusters of
non-luminous bodies. In 4-dimensional general relativity, 
a black hole is completely specified by its mass $M$ and 
by its spin parameter $a_*$, and it is subjected to the Kerr 
bound $|a_*|\le1$, which is the condition for the existence 
of the event horizon. The accretion process can spin a 
black hole up to $a_* \approx 0.998$ and at least some of 
the super-massive objects at the center of galaxies may be 
fast-rotating black holes with spin parameter close to this 
value.

The Kerr black hole paradigm is still based on a set of 
unproven assumptions. They sounded reasonable forty years
ago, but they are more questionable today. There is no 
direct evidence that black hole candidates have an 
event horizon, while there 
are theoretical arguments suggesting new physics appearing 
at macroscopic scales~\cite{np1,np2,np3,np4,np5,np6}. If 
the astrophysical black hole candidates are not 
the objects predicted by general relativity, they are not 
subjected to the Kerr bound. Interestingly, the accretion 
process onto a body with non-Kerr quadrupole moment typically 
spins the object up to $a_*>1$.

In Ref.~\cite{ss}, I used the Manko--Novikov solution to
describe the exterior gravitational field of a generic compact 
body. I showed that in most cases the equilibrium spin parameter
is larger than 1. However, the Manko--Novikov solution is
valid only for $|a_*|<1$ and therefore it was impossible to 
study the accretion process for $a_*>1$ and figure out the 
properties of the space-time around a super-spinning body. In 
the present paper, I considered the Manko--Mielke--Sanabria-G\'omez 
solution, which can describe the gravitational field outside
a compact body with non-Kerr quadrupole moment and spin parameter
either smaller and larger than 1. I studied the basic properties 
of the space-times when $a_*>1$, I discussed the evolution of 
the spin parameter, and I found its equilibrium value, 
see Figs.~\ref{f-4-1}, \ref{f-4-3}, and Tab.~\ref{t-eq}. 
For fast-rotating objects, the accretion processes onto a black 
hole and onto a generic body are quite different. For example, 
only around a black hole can the inner radius of the disk be 
very small, while the angular frequency of the gas particles at 
the innermost circular orbit can be significantly higher than 
the one around another object.

The fact that the accretion process in a non-Kerr background
can spin the compact body up to $a_* > 1$ is relevant for
the research devoted to figure out how future observations
can test the Kerr nature of the current astrophysical 
black hole candidates. The possibility that these objects
can have spin parameter larger than 1 cannot be ignored and
this is particularly true for experiments like LISA, whose 
detection relies on matched filtering.


\acknowledgments
I would like to thank Enrico Barausse, Shinji Mukohyama, and Naoki 
Yoshida for useful discussions. This work was supported by World 
Premier International Research Center Initiative (WPI Initiative), 
MEXT, Japan, and by the JSPS Grant-in-Aid for Young Scientists 
(B) No. 22740147.


\appendix

\section{Coordinate systems \label{a-1}}

The canonical form of the line element of a generic stationary and
axisymmetric space-time in quasi-cylindrical coordinates $\rho z$ is
\be\label{eq-ds}
ds^2 = - f \left(dt - \omega d\phi\right)^2
+ \frac{e^{2\gamma}}{f}\left(d\rho^2 + dz^2\right)
+ \frac{\rho^2}{f} d\phi^2 \, .
\ee
The relation between quasi-cylindrical coordinates $\rho z$ and 
prolate spheroidal coordinates $xy$ is qiven by
\be
\rho = k \sqrt{(x^2 - 1)(1 - y^2)} \, , \quad
z = kxy \, ,
\ee
where $k$ is a constant. The inverse relation is
\be
x = \frac{R_+ + R_-}{2k} \, , \quad
y = \frac{R_+ - R_-}{2k} \, ,
\ee
where $R_\pm = \sqrt{\rho^2 + (z \pm k)^2}$. Through the formal
transformation $x \rar ix$ and $k \rar -ik$, where $i$ is the
imaginary unit, one changes the prolate spheroidal coordinates 
into oblate spheroidal coordinates. The relation between the
quasi-cylindrical coordinates and the oblate spheroidal coordinates
is therefore given by
\be
\rho = k \sqrt{(x^2 + 1)(1 - y^2)} \, , \quad
z = kxy \, .
\ee
Lastly, the relation between quasi-cylindrical coordinates and
Schwarzschild coordinates is~\cite{glass}
\be\label{eq-rad}
\rho = \sqrt{r^2 - 2 M r + a^2} \sin\theta \, , \quad
z = (M - r) \cos\theta \, ,
\ee
where $M$ is the mass and $a=J/M$ the specific spin angular momentum.

\begin{figure}
\par
\begin{center}
\includegraphics[type=pdf,ext=.pdf,read=.pdf,width=7cm]{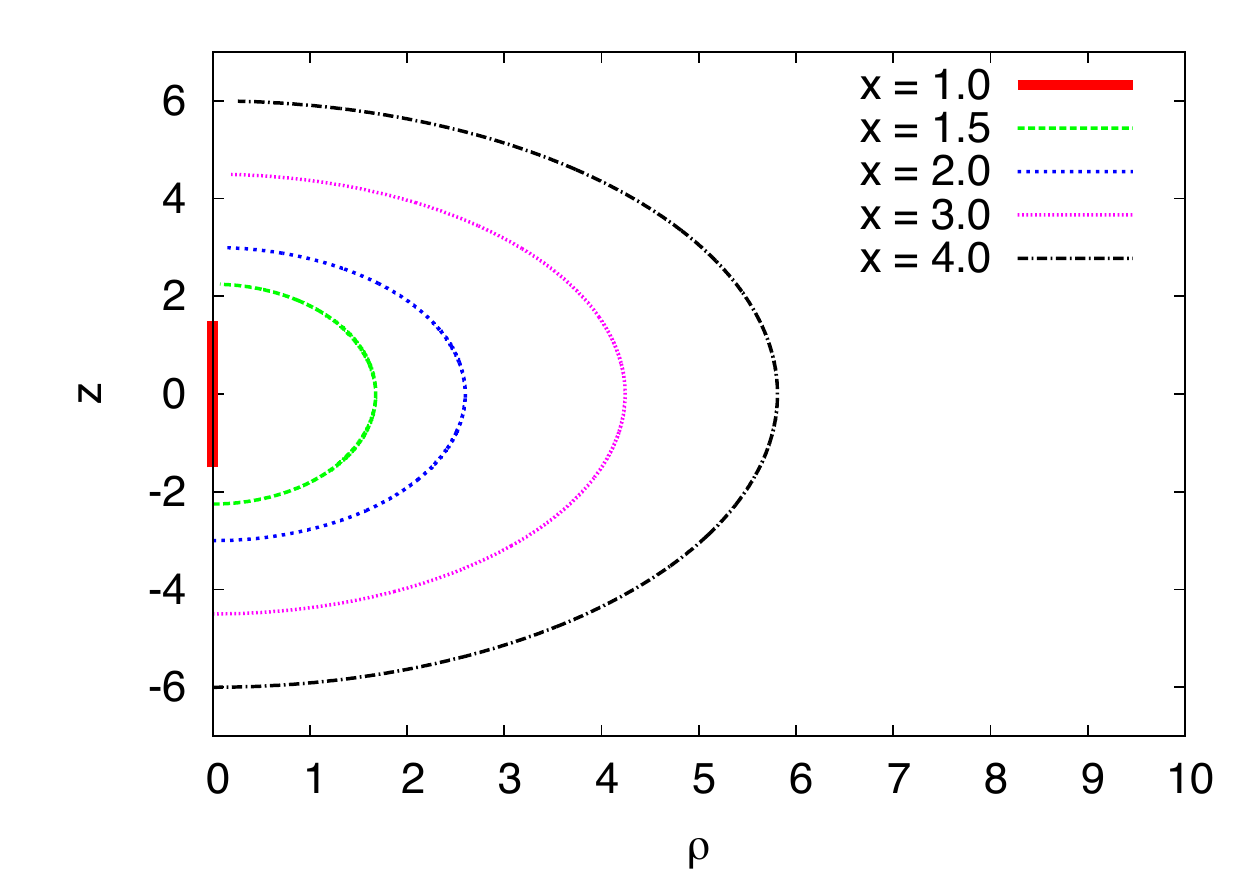}
\includegraphics[type=pdf,ext=.pdf,read=.pdf,width=7cm]{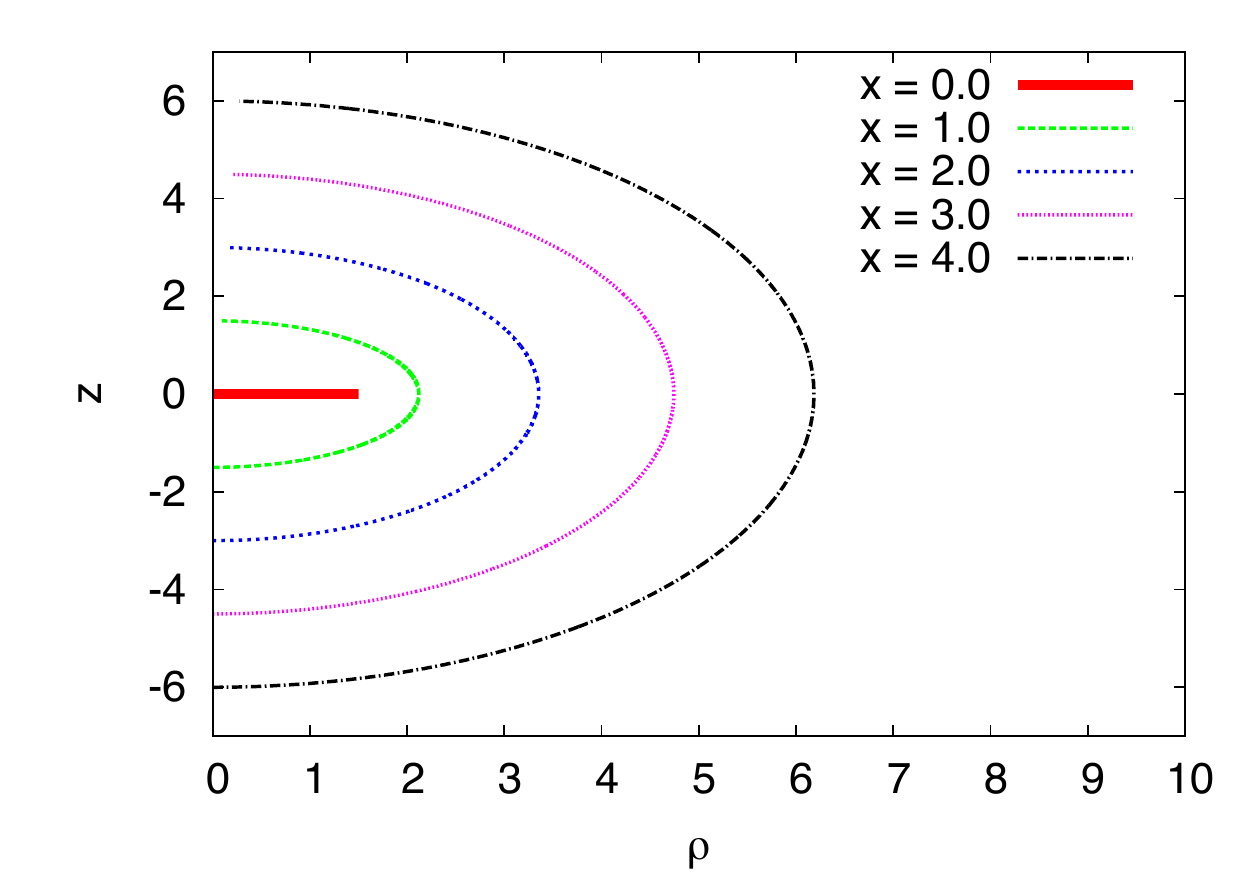}
\end{center}
\par
\vspace{-5mm} 
\caption{Curves with constant prolate spheroidal coordinate
$x$ (left panel) and oblate spheroidal coordinate $x$ (right
panel) on the $\rho z$-plane. $\rho$ and $z$ are given in units 
of $M=1$. $k=1.5 \, M$.}
\label{f-a-1a}
\par
\begin{center}
\includegraphics[type=pdf,ext=.pdf,read=.pdf,width=7cm]{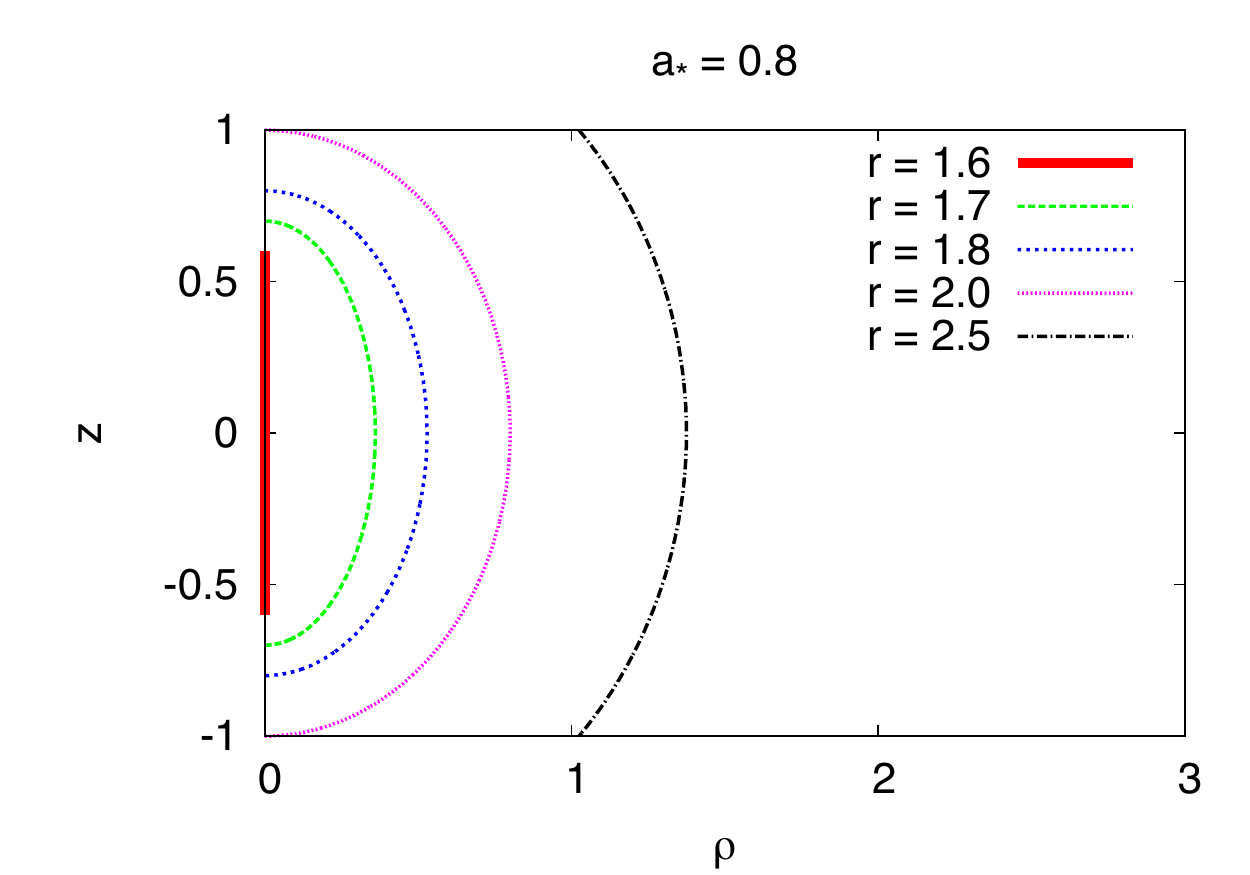}
\includegraphics[type=pdf,ext=.pdf,read=.pdf,width=7cm]{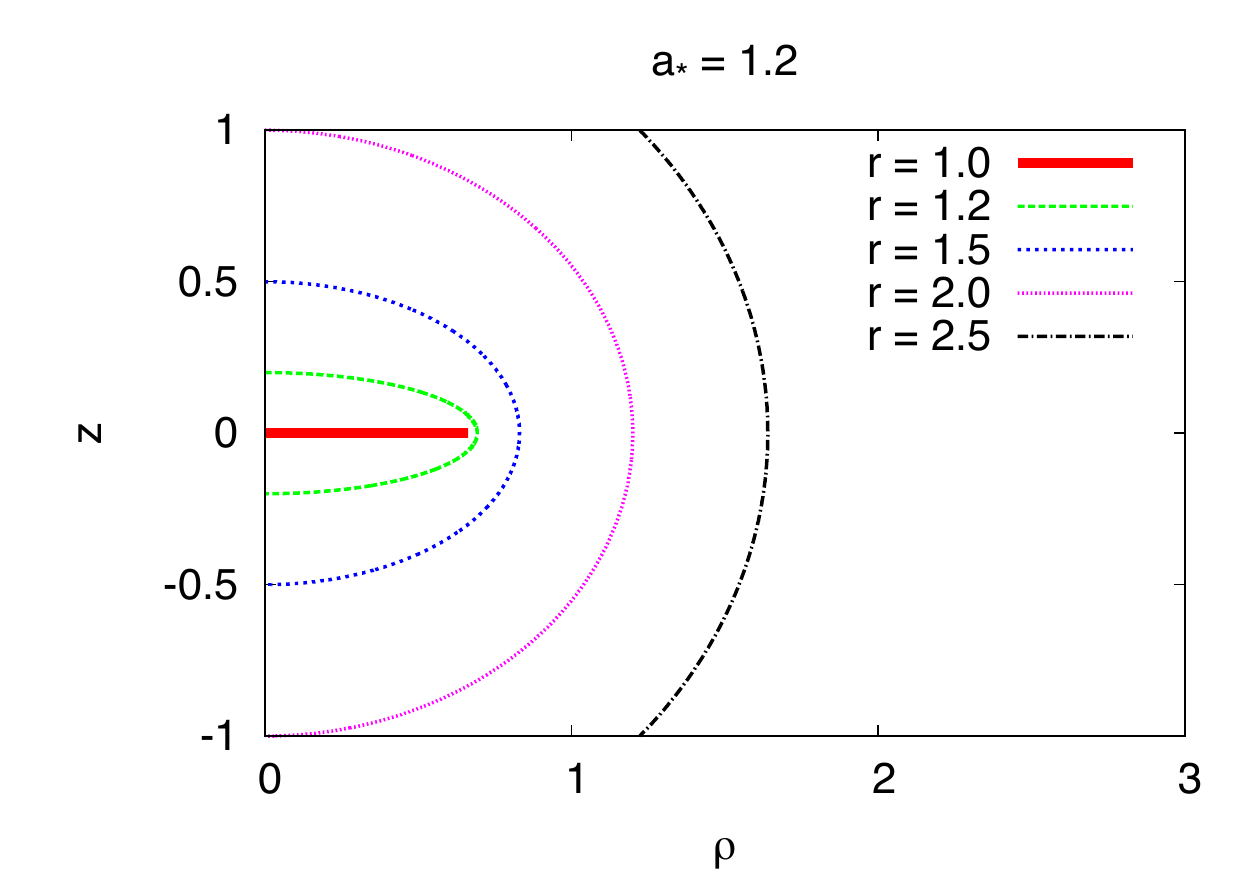}
\end{center}
\par
\vspace{-5mm} 
\caption{Curves with constant Schwarzschild radial coordinate
on the $\rho z$-plane, for $a_* =0.8$ (left panel) and 
$a_* = 1.2$ (right panel). $\rho$ and $z$ are given in units of $M=1$.}
\label{f-a-1b}
\end{figure}

Prolate spheroidal coordinates are suitable for describing the 
space-time around slow-rotating objects, while oblate spheroidal
coordinates are suitable in the case of fast-rotating objects. 
In the special case of the Kerr space-time, prolate spheroidal
coordinates can be used only for BHs, oblate spheroidal coordinates
only for Kerr naked singularity (see App.~\ref{a-2}).
The left (right) panel of Fig.~\ref{f-a-1a} shows some curves with 
constant prolate (oblate) coordinate $x$ on the $\rho z$-plane for 
$k=1.5 \, M$. Fig.~\ref{f-a-1b} shows instead some curves with constant
Schwarzschild radial coordinate, still on the $\rho z$-plane, for 
$a = 0.8 \, M$ (left panel) and $a = 1.2 \, M$ (right panel). In 
the former case, the quasi-cylindrical coordinates cover the space 
with Schwarzschild radial coordinate $r \ge r_H = M + \sqrt{M^2 
- a^2}$ and the surface with radial coordinate $r = r_H$ 
corresponds to the segment $\rho = 0$ and $|z| < \sqrt{M^2 - 
a^2}$. Here $r_H$ is equal to the Schwarzschild radial coordinate 
of the even horizon of a Kerr BH with spin $a$. When $|a| > M$, 
the quasi-cylindrical coordinates can be used to describe the 
region $r>M$ (otherwise there is not a one-to-one correspondence 
between the two coordinate systems) and the circle with radius 
$r=M$ reduces to the segment $|\rho|<\sqrt{a^2 - M^2}$ and $z=0$ 
on the $\rho z$-plane. For non-Kerr objects, prolate (oblate) 
spheroidal coordinates are still adequate only for slow-rotating 
(fast-rotating) objects, but in general the value of the spin 
$a$ separating the two cases is not $M$ any more.

\section{Kerr space-time in spheroidal coordinates\label{a-2}}

\begin{figure}
\par
\begin{center}
\includegraphics[type=pdf,ext=.pdf,read=.pdf,width=5cm]{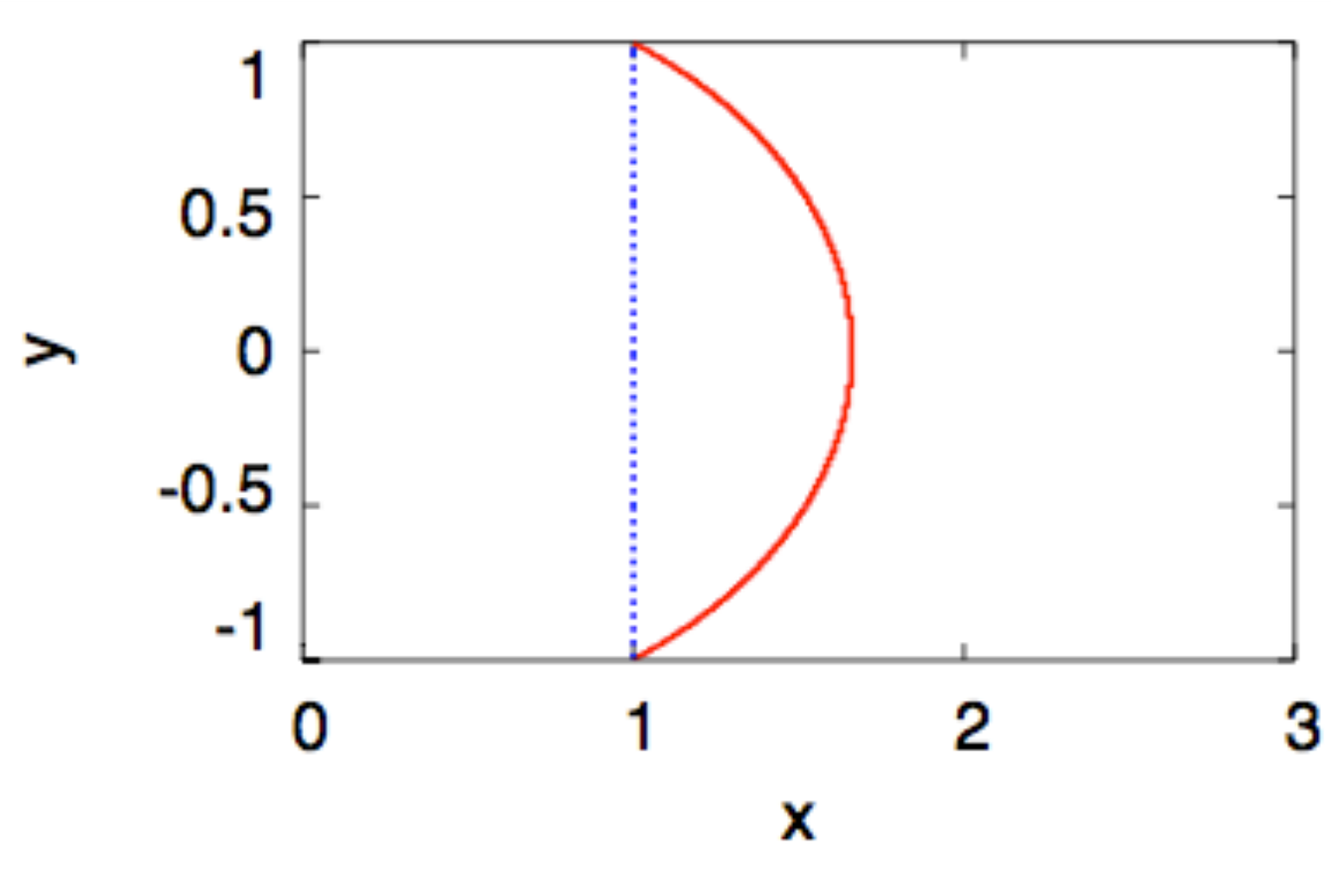}
\includegraphics[type=pdf,ext=.pdf,read=.pdf,width=5cm]{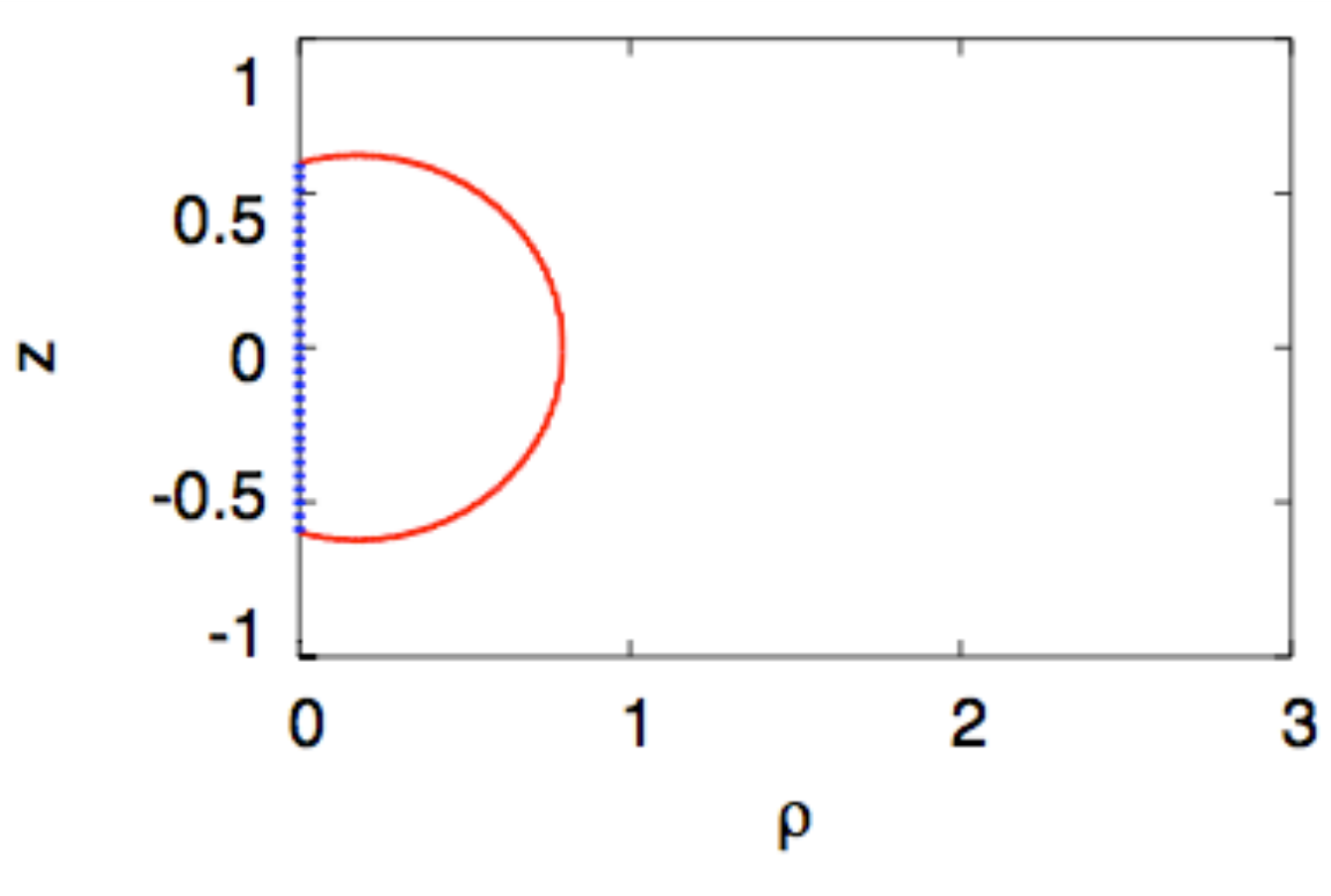}
\includegraphics[type=pdf,ext=.pdf,read=.pdf,width=5cm]{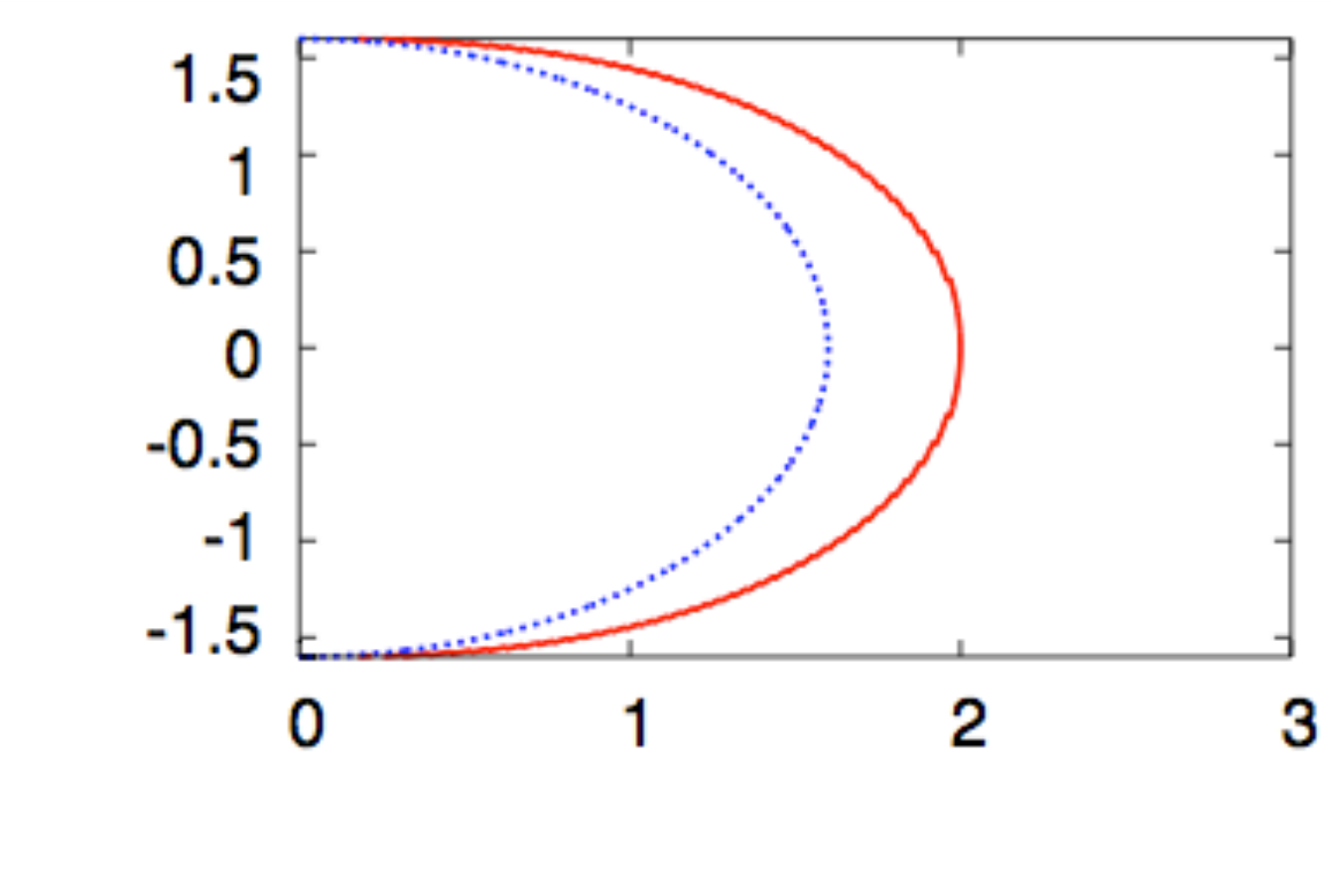}
\end{center}
\par
\vspace{-5mm} 
\caption{Infinite redshift surface $g_{tt}=0$ (red solid curves)
and event horizon (blue dotted curves) in Kerr space-time with 
$a_* = 0.8$ in prolate spheroidal coordinates $xy$ (left panel), 
quasi-cylindrical coordinates $\rho z$ (central panel), and 
Schwarzschild coordinates $r\theta$ (right panel). $\rho$, 
$z$, and $r$ are given in units of $M = 1$.}
\label{f-a-2}
\par
\begin{center}
\includegraphics[type=pdf,ext=.pdf,read=.pdf,width=5cm]{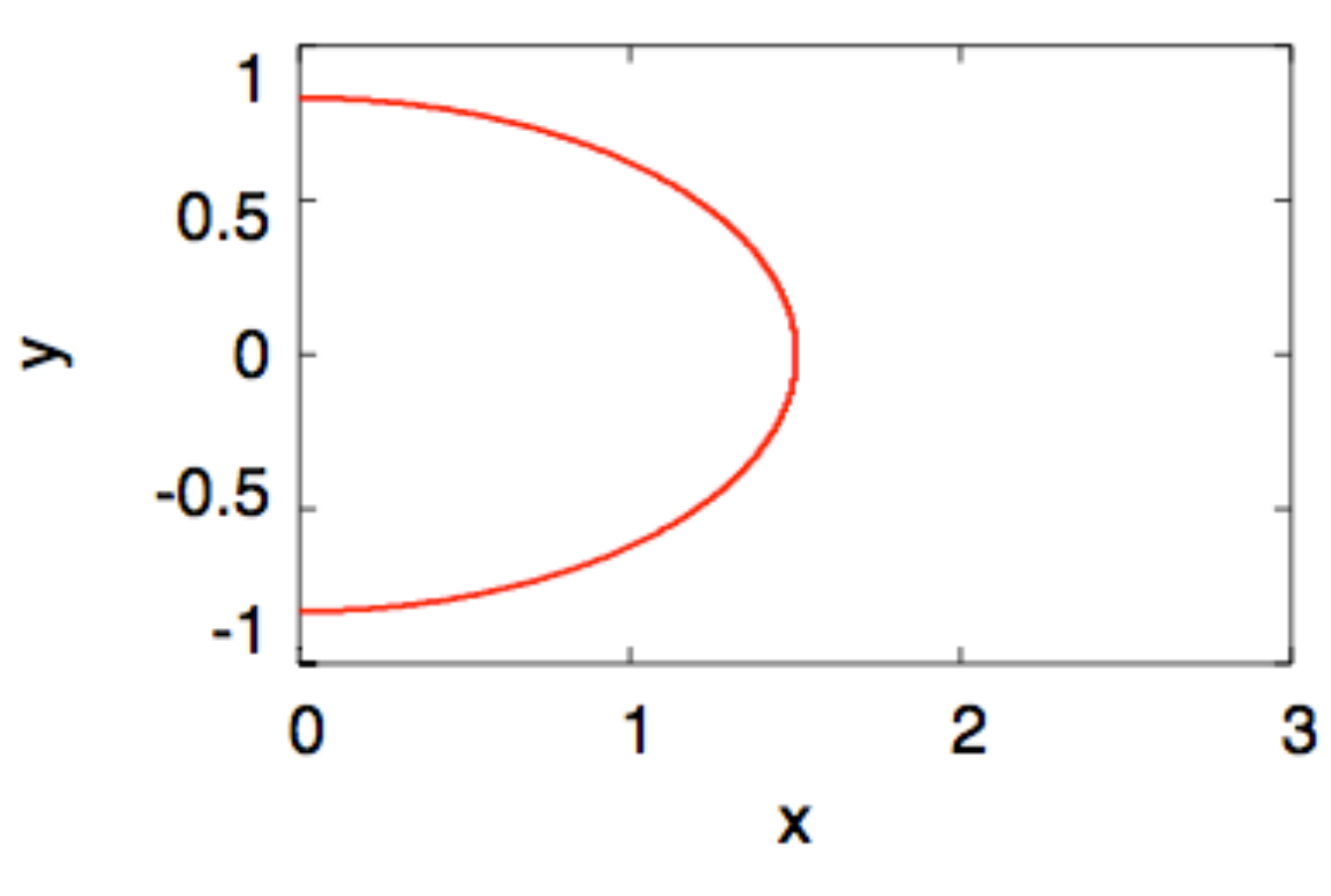}
\includegraphics[type=pdf,ext=.pdf,read=.pdf,width=5cm]{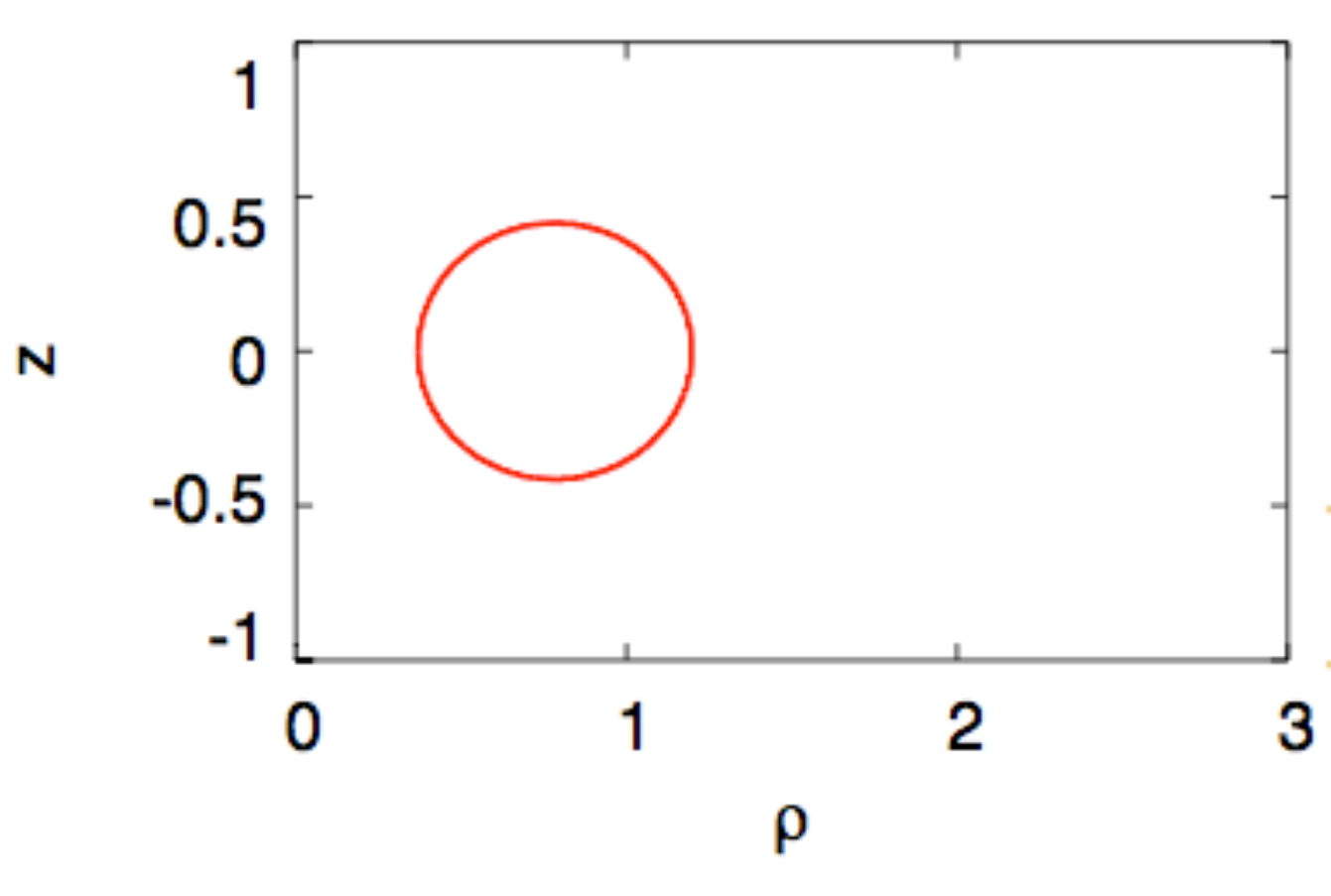}
\includegraphics[type=pdf,ext=.pdf,read=.pdf,width=5cm]{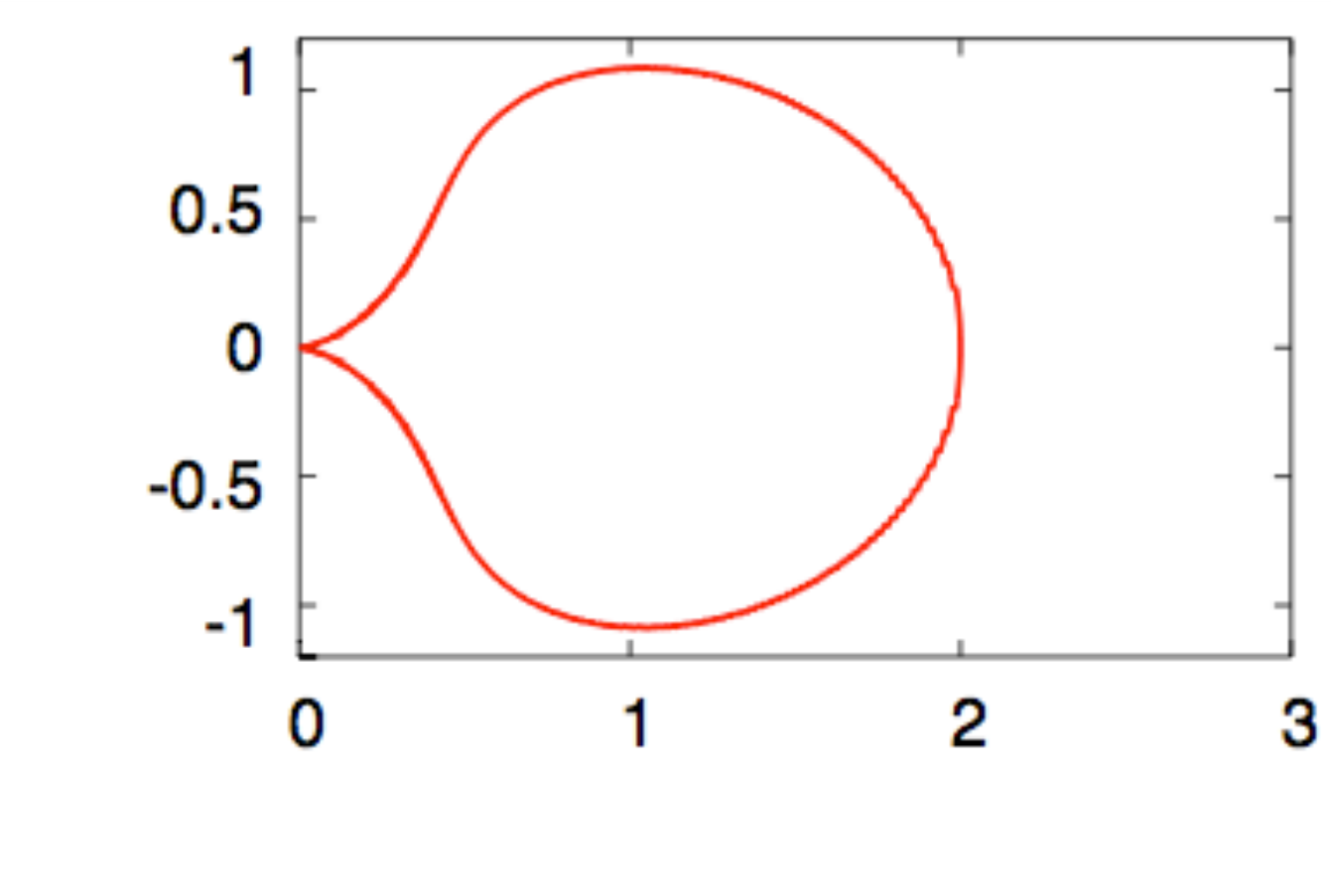}
\end{center}
\par
\vspace{-5mm} 
\caption{Infinite redshift surface $g_{tt}=0$ in Kerr space-time 
with $a_* = 1.2$ in oblate spheroidal coordinates $xy$ (left panel), 
quasi-cylindrical coordinates $\rho z$ (central panel), and 
Schwarzschild coordinates $r\theta$ (right panel). $\rho$, 
$z$, and $r$ are given in units of $M = 1$.}
\label{f-a-3}
\end{figure}

In prolate spheroidal coordinates, the line element~(\ref{eq-ds}) is
\be
ds^2 &=& - f \left(dt - \omega d\phi\right)^2
+ \frac{k^2 e^{2\gamma}}{f}\left(x^2 - y^2\right)
\left(\frac{dx^2}{x^2 - 1} + \frac{dy^2}{1 - y^2}\right)
+ \nonumber\\ &&
+ \frac{k^2}{f} \left(x^2 - 1\right)\left(1 - y^2\right) d\phi^2 \, .
\ee
In the Kerr space-time
\be
f = \frac{A}{B} \, , \quad 
\omega = - (1 - y^2) \frac{C}{A} \, , \quad
e^{2\gamma} = \frac{A}{k^2(x^2 - y^2)} \, ,
\ee
where $A$, $B$, and $C$ are given by 
\be
A = k^2 (x^2 - 1) - a^2 (1 - y^2) \, , \quad
B = (kx + M)^2 + a^2 y^2 \, , \quad
C = 2 a M (kx + M) \, .
\ee
Here $M$ is the mass, $a=J/M$ is the specific spin angular 
momentum, and $k = \sqrt{M^2 - a^2}$. Prolate 
spheroidal coordinates can be used to describe only BHs 
($|a_*| < 1$) and the surface $x=1$ is the BH event horizon.
The transformation $x\rar ix$ and $k \rar -ik$ changes
the prolate spheroidal coordinates into oblate spheroidal
coordinates, which can be used to describe the space region
around a Kerr naked singularity ($|a_*| > 1$) with Schwarzschild
radial coordinate $r>M$. Fig.~\ref{f-a-2} shows the infinite
redshift surface $g_{tt}=0$ (red solid curve) and the event
horizon (blue dotted curve) of a BH with $a_*=0.8$ in 
{\it prolate} spheroidal coordinates $xy$ (left panel), 
quasi-cylindrical coordinates $\rho z$ (central pane), and 
Schwarzschild coordinates $r\theta$ (right panel).
Fig.~\ref{f-a-3} shows the infinite redshift surface of a 
Kerr naked singularity with $a_* = 1.2$ in {\it oblate} 
spheroidal coordinates $xy$ (left panel), quasi-cylindrical 
coordinates $\rho z$ (central pane), and Schwarzschild 
coordinates $r\theta$ (right panel).

\end{document}